\documentclass[conference]{IEEEtran}
\usepackage[colorlinks=true,allcolors=blue,breaklinks,draft=false]{hyperref}   % hyperlinks, including DOIs and URLs in bibliography
\usepackage[cmex10]{amsmath}
\usepackage{amsthm}
\usepackage{amssymb}

\usepackage{algpseudocode}
\algrenewcommand\algorithmicindent{0.5em}
\usepackage{tcolorbox}
\tcbset{left=0pt,right=5pt,top=5pt,bottom=5pt, boxrule=1pt}

\usepackage[caption=false,font=footnotesize]{subfig}

\usepackage{multicol}
\usepackage{lipsum}
\usepackage[export]{adjustbox}

\usepackage{listings}
\makeatletter
\lst@CCPutMacro\lst@ProcessOther {"2D}{\lst@ttfamily{-{}}{-{}}}
\@empty\z@\@empty
\makeatother
\usepackage{color}
\usepackage{textcomp}
\definecolor{lbcolor}{rgb}{0.93,0.93,0.93}
\newcommand{\bdescription}{\begin{description}}
\newcommand{\edescription}{\end{description}}

\begin{document}

%
% any author declaration will be ignored  when using 'plid' option (for double blind review)
%

\title{Fusion of Array Operations at Runtime}

\author{\IEEEauthorblockN{Mads R. B. Kristensen,
Simon A. F. Lund,
Troels Blum, and
James Avery}
\IEEEauthorblockA{Niels Bohr Institute, University of Copenhagen, Denmark\\
    \{madsbk/safl/blum/avery\}@nbi.ku.dk
}
}

\maketitle

\newcommand{\opname}[1]{\mathsf{#1}}

\newcommand{\cost}{\opname{cost}}
\newcommand{\MWCcost}{\opname{cost}_{\mathrm{MWC}}}
\newcommand{\cut}{\opname{cut}}
\newcommand{\setof}[2]{\left\{#1\middle| #2\right\}}
\newcommand{\bytecount}[1]{\left\|#1\right\|}
\newcommand{\bigO}[1]{\ensuremath{\mathcal{O}\left(#1\right)}}

\newcommand{\depop}[1]{%
  \mathrel{\vbox{\offinterlineskip\ialign{%
    \hfil##\hfil\cr
    $\scriptscriptstyle d$\cr
    \noalign{\kern0ex}
    $#1$\cr
}}}}
%JA: MRBK: Lad os lige finde ud af, hvordan vi paenest skriver det her.
\newcommand{\ltdep}{\depop{<}}          % Dependency order '<'
\newcommand{\ltedep}{\depop{\ge}}       % Dependency order '<='
\newcommand{\precdep}{\depop{\prec}}    % Dependency order predecessor
\newcommand{\succdep}{\depop{\succ}}    % Dependency order successor

\newcommand{\ltpar}{<}          % Partition order '<'
\newcommand{\ltepar}{\ge}       % Partition order '<='
\newcommand{\precpar}{\prec}    % Partition order predecessor
\newcommand{\succpar}{\succ}    % Partition order successor

\newcommand{\pgraph}[2]{\hat{#1}(#2)}
\newcommand{\pgraphEd}[1]{\pgraph{E_d}{#1}}
\newcommand{\pgraphEf}[1]{\pgraph{E_f}{#1}}
\newcommand{\pgraphEw}[1]{\pgraph{E_f}{#1}}
\newcommand{\mergeop}{/}

\newcommand{\inop}{\opname{in}}
\newcommand{\outop}{\opname{out}}
\newcommand{\newop}{\opname{new}}
\newcommand{\delop}{\opname{del}}
\newcommand{\extop}{\opname{ext}}
\newcommand{\savingop}{\opname{saving}}

\newtheorem{theorem}{Theorem}
\newtheorem{prop}{Proposition}
\newtheorem{definition}{Definition}
\newtheorem{corollary}{Corollary}
\newtheorem{lemma}{Lemma}
\newtheorem{remark}{Remark}

\newcommand{\argmin}{\operatornamewithlimits{argmin}}

%add a medium cup and cap symbol with limit
\newcommand*{\mmedcup}{\mathbin{\scalebox{1.6}{\ensuremath{\cup}}}}
\newcommand*{\mmedcap}{\mathbin{\scalebox{1.6}{\ensuremath{\cap}}}}
\newcommand{\medcup}{\operatornamewithlimits{\mmedcup}}
\newcommand{\medcap}{\operatornamewithlimits{\mmedcap}}

%add a larger vbar "|"
\newcommand*{\vbar}{\mathbin{\scalebox{1.2}{\ensuremath{|}}}}

% Make sure that either the Long or the Short version flag is set
% If none of them is set, we set the LongVersion flag as default
\ifdefined\ShortVersion
    \ifdefined\LongVersion
        \PackageError{Macros}{Both long and short version macro is defined!}{}
    \fi
    \PackageInfo{Macros}{Building the Short version}
\else
    \ifdefined\LongVersion
        \newcommand{\ShortVersion}{}
    \else
        \newcommand{\LongVersion}{}
    \fi
    \PackageInfo{Macros}{Building the Long version (default)}
\fi

\begin{abstract}
    We address the problem of fusing array operations based on criteria
such as shape compatibility, data reusability, and communication. We
formulate the problem as a graph partition problem that is general
enough to handle loop fusion, combinator fusion, and other types of
subroutines.

\end{abstract}

\newsavebox{\LstCFuseExampleA}
\begin{lrbox}{\LstCFuseExampleA}
\begin{lstlisting}[language=c, numbers=none, linewidth=0.41\linewidth]
#define N 1000
double A[N], B[N], T[N];
for(int i=0; i<N; ++i)
  T[i] = B[i] * A[i];
for(int i=0; i<N; ++i)
  A[i] += T[i];
\end{lstlisting}
\end{lrbox}

\newsavebox{\LstCFuseExampleB}
\begin{lrbox}{\LstCFuseExampleB}
\begin{lstlisting}[language=c, numbers=none, linewidth=0.41\linewidth]
for(int i=0; i<N; ++i){
  T[i] = B[i] * A[i];
  A[i] += T[i];
}
\end{lstlisting}
\end{lrbox}

\newsavebox{\LstCFuseExampleC}
\begin{lrbox}{\LstCFuseExampleC}
\begin{lstlisting}[language=c, numbers=none, linewidth=0.41\linewidth]
for(int i=0; i<N; ++i){
  double t = B[i] * A[i];
  A[i] += t;
}
\end{lstlisting}
\end{lrbox}

\newsavebox{\LstCNonFuseExample}
\begin{lrbox}{\LstCNonFuseExample}
\begin{lstlisting}[language=c, numbers=none, linewidth=0.41\linewidth]
#define N 1000
double A[N], B[N], T[N];
int j = N;
for(int i=0; i<N; ++i)
  T[i] = B[i] * A[i];
for(int i=0; i<N; ++i)
  A[i] += T[--j];
\end{lstlisting}
\end{lrbox}

\section{Introduction}

Array operation fusion is a program transformation that combines, or
fuses, multiple array operations into a \emph{kernel} of
operations. When it is applicable, the technique can drastically
improve cache utilization through temporal data locality and enables
other program transformations such as streaming and array
contraction~\cite{Gao93_array_contraction}. In scalar programming
languages, such as C, array operation fusion typically corresponds to
loop fusion where multiple computation loops are combined into
a single loop. The effect is a reduction of array traversals
(Fig. \ref{lst:fuse_C_code}). Similarly, in functional
programming languages it typically corresponds to fusing
individual combinators. In array programming languages, such as
HPF~\cite{loveman1993high} and ZPL~\cite{zpl00}, fusing array operations
are crucial, since a program written in these languages will consist
almost exclusively of array operations. Lewis et al. demonstrates a
execution time improvement of up to 400\% when optimizing for array
contraction at the array rather than the loop level~\cite{Lewis1998}.

However, not all fusions of operations are
allowed. Consider the two loops in
Fig. \ref{lst:nonfuse_example_in_c}; since the second loop traverses
the result from the first loop in reverse, we must compute the
complete result of the first loop before continuing to the second loop,
preventing fusion.
Clever analysis sometimes allows transforming the program into a form that is amenable to fusion, but
such analysis is outside the scope of the present work.
Throughout the remainder of this paper, we assume that any such optimizations have already been performed.

Deciding which operations to fuse together is the same as
finding a partition of the set of operations in which the blocks obey
the same execution dependency order as the individual operations, and in
which no block contains two operations that may not be
fused. Out of all such partitions, we want to find one that enables us to
save the most computation or memory.  It is not an easy problem, in
part because fusibility is not transitive. That is, even when it is
legal to fuse subroutines $x,y$ and $y,z$, it may be illegal for
all three of $x,y,z$ to be executed together. Thus, one local decision
can have global consequences on future possible partitions.

The problem can be stated in a quite general way: ``\emph{Given a
  mixed graph, find a legal partition of vertices that cuts all
  non-directed edges and minimizes the cost of the
  partition.}''\footnote{See Sec. \ref{sec:subroutine_partition} for
  the definition of a legal partition and legal cost function.}. We
call this problem the \emph{Weighted Subroutine Partition} problem,
abbreviated WSP.

\begin{figure}
    \centering
    \subfloat[][Two forward iterating loops.]{\label{lst:fuse_C_code_a}\usebox{\LstCFuseExampleA}} \hspace{40px}
    \subfloat[][A forward and a reverse iterating loop.]{\label{lst:nonfuse_example_in_c}\usebox{\LstCNonFuseExample}}\\
    \subfloat[][Loop fusion: the two loops from Fig. \ref{lst:fuse_C_code_a} fused into one.]{\label{lst:fuse_C_code_b}\usebox{\LstCFuseExampleB}}\hspace{40px}
    \subfloat[][Array contraction: the temporary array \texttt{T} from Fig. \ref{lst:fuse_C_code_b} is contracted into the scalar \texttt{t}.]{\label{lst:fuse_C_code_c}\usebox{\LstCFuseExampleC}}
    \caption{Loop fusion and array contraction in C.}
    \label{lst:fuse_C_code}
\end{figure}

The general formulation is applicable to a broad range of optimization
objectives. The cost function can penalize any
aspect of the partitions, e.g.~data accesses, memory use,
communication bandwidth, and/or communication latency. The only
requirement to the cost function is monotonicity:
\begin{itemize}
\item Everything else equal, it must be cost neutral or a cost
  advantage to place two subroutines within the same partition block.
\end{itemize}
Similarly, the definition of partition legality is flexible.
\begin{itemize}
\item Any aspect of a pair of subroutines can make them illegal to
  have in the same partition block, such as preventing mixing of
  sequential and parallel loops, different array shapes, or access
  patterns.
\item Subroutines may have dependencies that impose a partial order.
  Then a legal partition must observe this order, i.e.~must not introduce cycles.
  % Then,
  % given three subroutines, $x \ltdep y \ltdep z$, where $x,z$ are
  % in partition block $B$ then $y$ must also be in $B$ in order for
  % the partition to be legal.
\end{itemize}

The remainder of the paper is structured as follows: In Section 3, we formally define
the {\em Weighted Subroutine Partition problem}, which unifies array
operation-, loop-, and combinator fusion, and prove that it is
NP-hard. Section 4 shows how WSP is used to solve array operation
fusion for the {\em Bohrium} automatic parallelization framework, and gives a
correctness proof. In Section 5, we describe a branch-and-bound
algorithm that computes an optimal solution, as well as two approximation
algorithms that compute good results rapidly enough to use in
JIT-compilation. All the algorithms are implemented in Bohrium, and
work for any choice of monotonic cost function, which allows us to compare
directly with other fusion schemes from the literature.  Section 6
shows measurements performed on 15 benchmark programs, comparing both
the optimal to the approximation schemes, and to three other fusion
schemes.

% Key contributions of this paper:
% \begin{itemize}
%    \item Introduction of the WSP problem that unifies array operation, loop, and combinator fusion by incorporating the cost function into the theoretical formulation.
%    \item We prove that the WSP problem correctly solves the fusion of array operations problem, and that it is NP-hard.
%    \item We show that existing methods that represent cost as static edge weights are not sufficient to measure data access costs correctly. It is necessary to assign
%      costs directly to partitions.
%    \item We have implemented a branch-and-bound algorithm that finds the optimal solution to the WSP problem and works with any legal cost function,
%      as well as two approximation algorithms that are fast enough for Just-In-Time compilation and that give good but suboptimal solutions.
%    \item Efficiency of the generated code is measured on 15 benchmarks
% \end{itemize}

\section{Related Work}

The WSP problem presented in this paper generalizes %is a further development of
the \emph{Weighted Loop Fusion} (WLF) problem first described by
Kennedy in \cite{Kennedy1994} (by the name \emph{Fusion for
  Reuse}). The method aims
to maximize data locality through loop fusion (corresponding to the
\textbf{Max Locality} cost model in Section \ref{sec:alternative-cost}).
\ifdefined\LongVersion The WLF
problem is described as a graph problem where vertices represent
computation loops, directed edges represent data dependencies between
loops, and undirected edges represent data sharing between
loops. Edges that connect fusible loops have a non-negative weight
that represents the cost saving associated with fusion of the two
loops. Edges that connect non-fusible loops are marked as
fuse-preventing. Now, the objective is to find a partition of the
vertices into blocks such that no block has vertices connected with
fuse-preventing edges and that minimize the weight sum of edges that
connects vertices in different blocks.  \else
The WLF problem is described as a graph-partitioning problem on a
\emph{Loop Dependece Graph} with static weights between fusible loops.
\fi

Megiddo et al.~have shown that it is possible to formulate the WLF problem
as integer linear programming (ILP)~\cite{Megiddo1997}.
\ifdefined\LongVersion Based on the WLF graph, the idea is to
transform the edges into linear constraints that implement the
dependency and fusibility between the vertices and transform weights
into ILP objective variables. The values of the objective variables are
either the values of the weights when the associated vertices are in
different partitions or zero when in the same partition. The objective of
the ILP is then to minimize the value of the objective variables .  The
problem is NP-hard, but the hope is that with an efficient ILP solver,
such as \texttt{lp-solve}~\cite{lpsolve}, and a modest problem size it
might be practical as a compile time optimization.  \fi
Darte et al.~\cite{Darte2002} proved that the WLF problem is NP-hard
through a reduction from multiway cut \cite{dahlhaus1992}. Furthermore, since maximizing data
locality may not maximize the number of array
contractions\ifdefined\LongVersion (Fig. \ref{lst:alain_example})\fi,
they introduce an ILP formulation with the sole objective of maximizing
the number of array contractions (the \textbf{Max
  Contract} cost model in Section~\ref{sec:alternative-cost}).

Robinson et al. \cite{Robinson2014} describe an ILP scheme that
combines the objectives of Megiddo and Darte: both
maximizing data locality and array contractions while giving priority to data
locality (corresponds to our \textbf{Robinson} cost model).

However, optimization using WLF has a significant limitation: it
only allows static edge weights. That is, when building the WLF graph the
values of edge weights are assigned once and for all. This limitation
is the main reason that we needed to develop the Weighted Subroutine Partition
formalism, because static edge weights are in fact inappropriate for
accurate measurement of data locality.

\begin{figure*}
    \centering
    \subfloat[][]{\includegraphics[trim={10px 20px 10px 10px}, clip, width=0.5\linewidth]{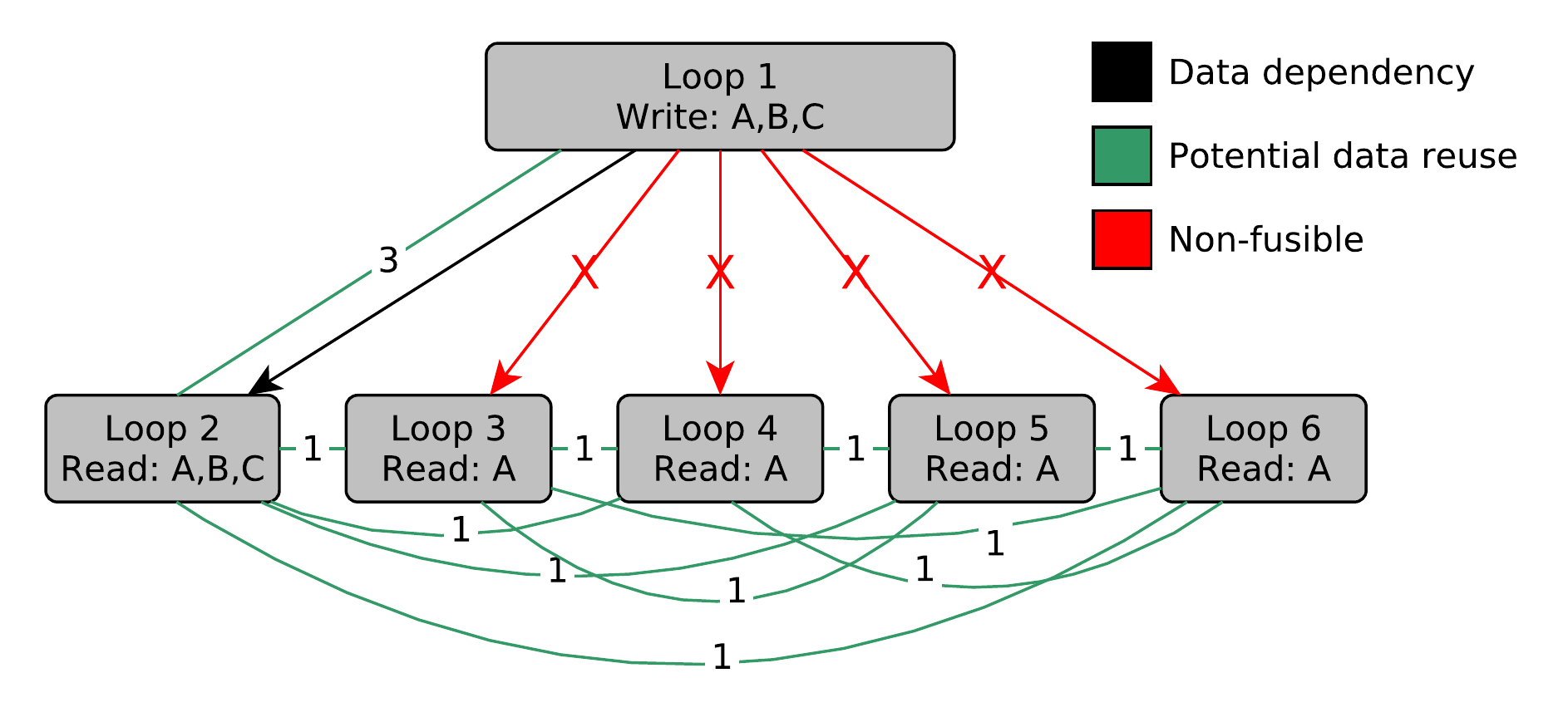}}\\
    \subfloat[][]{\label{fig:weighted_loop_fusion_graphB}\includegraphics[trim={15px 15px 10px 15px}, clip, width=0.5\linewidth]{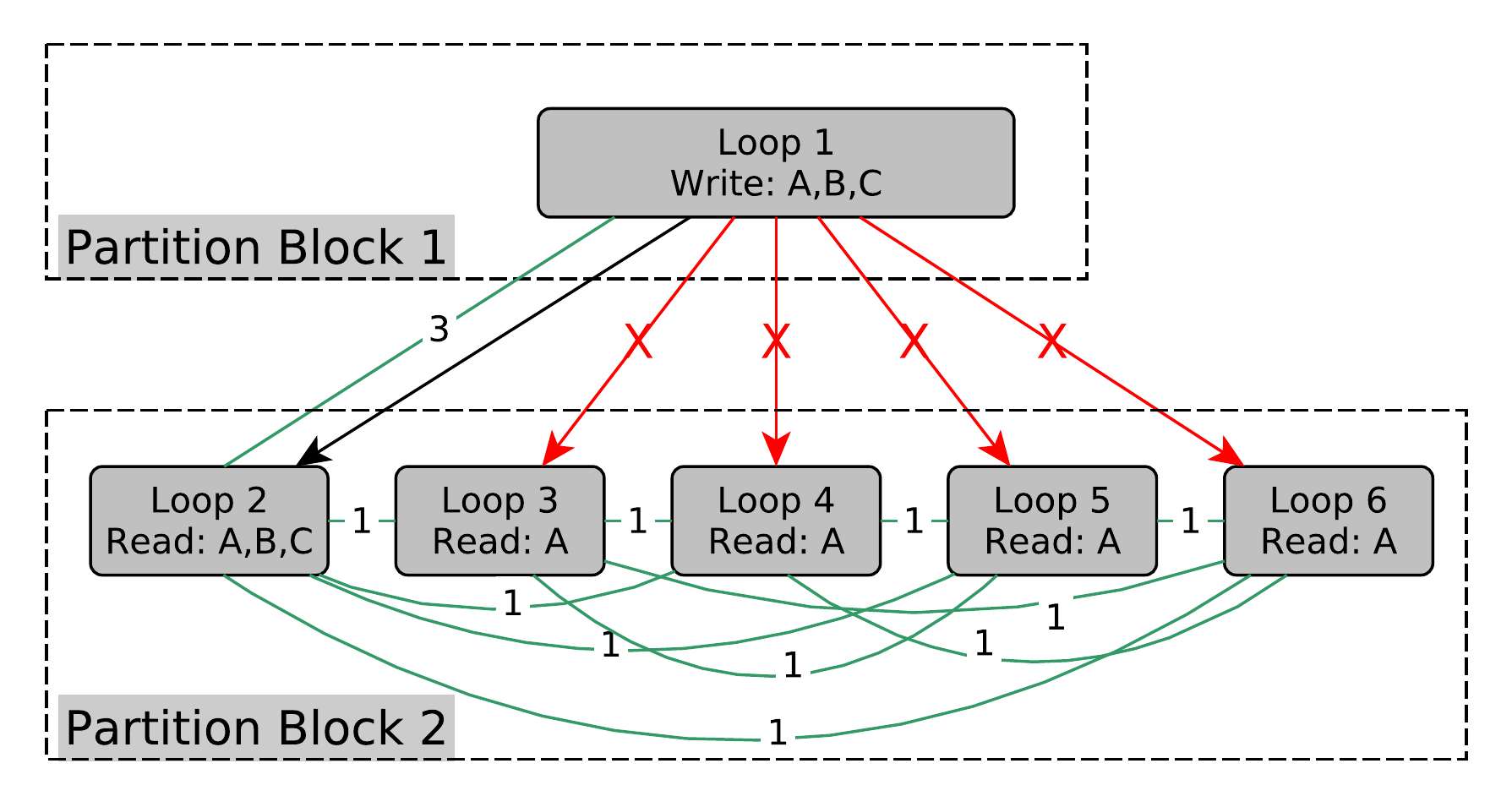}}
    \subfloat[][]{\label{fig:weighted_loop_fusion_graphC}\includegraphics[trim={10px 15px 15px 15px}, clip, width=0.5\linewidth]{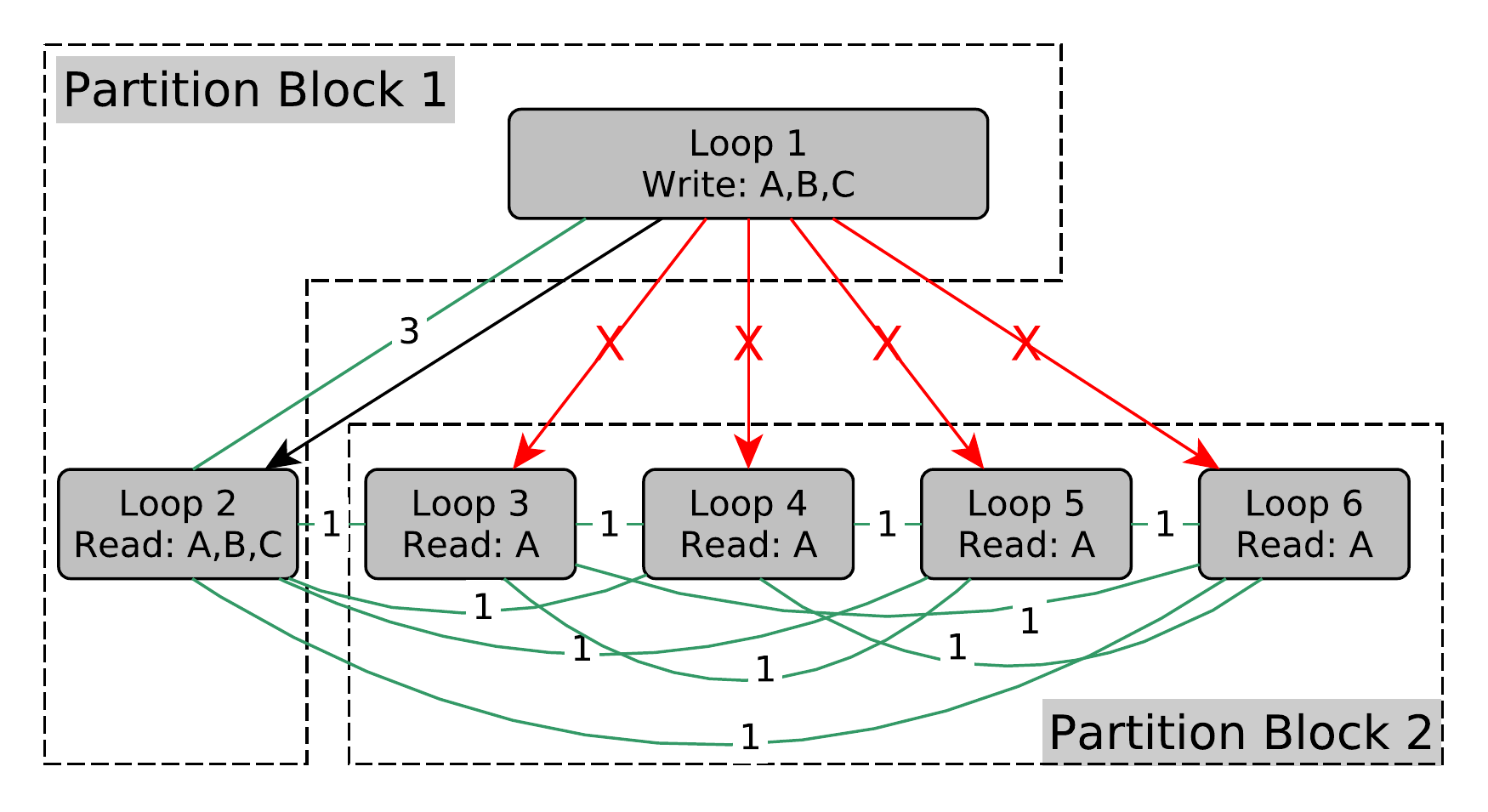}}
    \caption{A WLF example where the objective is to maximize data locality. (a) shows the initial graph. (b) shows a partition where loop 1 is in one block and loops 2-6 are in another block. (c) shows a partition where loops 1-2 are in one block and loops 3-6 are in another block.}
\label{fig:weighted_loop_fusion_graph}
\end{figure*}

Consider the WLF example in Fig.~\ref{fig:weighted_loop_fusion_graph},
which consist of six loops and three arrays $A,B,C$ of size $1$. The
objective is to maximize data locality, represented by weight edges connecting the
loops that access the same
arrays. Fig.~\ref{fig:weighted_loop_fusion_graphB} shows the optimal WLF
solution to the example, which reduces the total weight from 13 to
3. However, the actual number of array accesses is only reduced from
10 to 7. A better strategy is to fuse loop 1-2
(Fig.~\ref{fig:weighted_loop_fusion_graphC}), which will reduce the
actual number of array accesses from 10 to 4.

The problem with the WLF formulation here is that all the loops that read
the same data must be pair-wise connected with a weight, leading to
over-estimating potential data reuse. In the Weighted Subroutine
Partition formulation, we work with {\em partitions} instead of
individual merges, and assign a cost to a partition as a whole. The
cost-savings of a merge is then the difference in cost between the
partitions before and after merging, allowing accurate descriptions of
data-reuse through the costs function.
% Both WLF and WSP are NP hard, so in both cases, approximation
% algorithms must be used for the methods to be practical.

\section{The Weighted Subroutine Partition Problem}
\label{sec:subroutine_partition}

The Weighted Subroutine Partition (WSP) problem is an extension of the
\emph{The Weighted Loop Fusion Problem}~\cite{Kennedy1994} where we
include the weight function in the problem formulation. In this
section, we will formally define the WSP problem and show that it is
NP-hard.

\begin{definition}[WSP graph] %JA: Flyttet acyklicitet til definition af WSP-graf.
\label{def:wsp_graph}
A {\em WSP graph} is a triplet $G = (V, E_d, E_f)$ such that $(V, E_d)$ is a
directed acyclic graph describing dependency order, and $(V,E_f)$ is
an undirected graph of forbidden edges.
\end{definition}

\begin{definition}[WSP order]
\label{def:wsp_vertex_order}
A WSP graph, $G = (V, E_d, E_f)$, induces a partial order $\ltdep$ on
$V$ as follows: $v \ltdep v'$ iff there exists a path from $v$ to $v'$
in $(V,E_d)$. Since $(V,E_d)$ is acyclic, this partial order is
strict.
\end{definition}

\begin{definition}[Partitions]
\label{def:waof_partition}
A {\em partition} of a set $V$ is a set $P = \{B_1,B_2,\ldots,B_k\}$ such that $V$ is the disjoint union of the {\em blocks} $B_1,\ldots,B_k$.
The set $\Pi_V$ of all partitions of $V$
is partially ordered as $P \le P'$ iff $\forall B\in P\exists B' \in P'\colon B\subseteq B'$,
i.e.~if each block in $P$ is a subset of a block in $P'$.
\end{definition}
The set of partitions $\Pi_V$ is a lattice with bottom and top elements $\bot = \{V\}$ and $\top = \{\{V\}\}$.
The successors to a partition $P$ in the partition order are those partitions that are identical to $P$ except for merging two of the blocks.
Conversely, splitting a block results in a predecessor. We write $P \precpar P'$ if $P'$ is a successor to $P$. This defines a
binary {\em merge operator}:
\begin{definition}[Block merge operator]
  \label{def:block_merge}
  Given a partition $P = \{B_1,B_2,\ldots\}$, define $P \mergeop
  (B_1,B_2) = \{B_1\cup B_2,\ldots\}$ to be the successor to $P$ in
  which $B_1$ and $B_2$ are merged and all other blocks are left the
  same.
\end{definition}

\begin{definition}[Legal partition]
\label{def:wsp_legal_partition}
Given a WSP graph, $G=(V, E_d, E_f)$, we say that the partition $P\in
\Pi_V$ is {\em legal} when the following holds for every block $B\in P$:
\begin{enumerate}
    \item $\nexists v_1,v_2\in B: (v_1,v_2) \in E_f$, i.e.~ no block contains both endpoints of a forbidden edge.
    \item If $v_1 \ltdep v_2 \ltdep v_3$ and $v_1,v_3\in B$ then $v_2\in B$,  i.e.~ the directed edges between blocks must not form cycles.
\end{enumerate}
\end{definition}

\begin{definition}[WSP cost]
\label{def:wsp_cost}
Given a partition, $P$, of vertices in a WSP graph, a {\em cost function}
$\cost(P)$ returns the cost of the partition and respects the following
conditions:
\begin{enumerate}
    \item $\cost(P) \geq 0$
    \item $P \leq P' \implies \cost(P) \geq \cost(P')$
\end{enumerate}
\end{definition}

\begin{definition}[WSP problem]
\label{def:wsp_solution}
Given a WSP graph, $G = (V, E_d, E_f)$, and a cost function,
$\cost(P)$, the WSP problem is the problem of finding a legal
partition, $P^*$, of $V$ with minimal cost:%\vspace{-1.5em}
\begin{equation}
  \label{eq:wsp_solution}
    P^*\in \argmin_{P\in \hat{\Pi}_V}\cost(P)
\end{equation}
where $\hat{\Pi}_V$ denotes the set of legal partitions of $V$.
\end{definition}

%JA: Skal "complexity" egentlig være først?
\subsection{Complexity}
In order to prove that the WSP problem is NP-hard, we perform a
reduction from the \emph{Multiway Cut Problem}~\cite{dahlhaus1992},
which Dahlhaus et al.~has shown is NP-hard for fixed $k \geq 3$.

%JA: Farligt, hvis vi skal bruge MWC til at vise noget on WSP. Er det aekvivalent? I saa fald skal vi sige det.
%    Jeg proever lige med en omformulering.
%In order to match our needs, we define an instance of the Multiway Cut (MWC) problem a bit differently than Dahlhaus et al.:
\begin{definition}[Multiway Cut]
  \label{def:multiway_cut1}
  Given a tuple $(V,E,S,w)$ consisting of a graph $(V,E)$, a {\em terminal set} $S=\{s_1,\ldots,s_k\}$ of vertices,
  and a non-negative weight $w(u,v)$ for each edge $(u,v)\in E$, a {\em multiway cut} is an edge set $E'$
  the removal of which leaves each terminal in separate components. The solutions to the MWC problem are the multiway cuts
  of minimal total weight.
\end{definition}
% We reformulate the standard definition of MWC in terms of partitions to allow easier comparison to WSP:
% \begin{definition}[Multiway Cut 2]
%   \label{def:multiway_cut2}
%   Given a tuple $(V,E,S,w)$ consisting of a graph $(V,E)$, a {\em terminal set} $S=\{s_1,\ldots,s_k\}$ of vertices,
%   and a non-negative weight $w(u,v)$ for each edge $(u,v)\in E$, a {\em multiway cut} is a partition $P$ of $V$
%   such that no two terminals are contained in the same block of $P$.
%   The cost of a partition $P$ is
%   \[
%    \MWCcost(P) = \sum_{\substack{B,B'\in P \\ B \neq B'}}\sum_{\substack{u \in B\\ v \in B' \\ (u,v)\in E}}w(u,v)
%   \]
%   We write $\hat{\Pi}_V$ for the set of partitions that are multiway cuts.
%   Then the solutions to the MWC problem are
%   \[
%      P^* \in \argmin_{P\in \hat{\Pi}_V} \MWCcost(P)
%   \]
% \end{definition}

\begin{theorem}
\label{tho:waog_is_np}
    The WSP problem is NP-hard for graphs $G = (V, E_d, E_f)$
    that have a chain of three or more edges in $E_f$.
\end{theorem}
\begin{proof}
We prove NP-hardness through a reduction from multiway cut.
Given an MWC-instance, $(V,E,S,w)$, we build a WSP-instance as follows. Let $G = (V,E_d,E_f)$,
$E_f = \{(s_i,s_j):1\leq i<j\leq k\}$, and $E_d = \emptyset$.
Define the {\em cut} of a partition as the set of edges that connect the blocks: \[\cut(P) = \setof{(u,v)\in E_f}{\nexists B\in P: (u,v)\in B}\]
The cuts of the legal WSP partitions $\hat{\Pi}_V$ are exactly the set of multiway cuts:
\begin{itemize}
\item The set of directed edges in $E_d$ is empty, which makes
  Def.~\ref{def:wsp_vertex_order} and
  Def.~\ref{def:wsp_legal_partition}(2) trivially satisfied.
\item The fuse-preventing edges $E_f$ connect each terminal
  in $S$ and no other vertices. Hence, by
  Def.~\ref{def:wsp_legal_partition}(1), $\hat{\Pi_V}$ are exactly
  those partitions for which no block contains two terminals.
\end{itemize}
Let now the cost function be the total weight of the cut:
\[
 \cost(P) = \sum_{(u,v)\in \cut(P)} w(u,v)
\]
This is a valid WSP cost function (by Def.~\ref{def:wsp_cost}): it
is non-negative, and if $P \le P'$ in the partition order, then
$\cut(P) \supseteq \cut(P')$, whereby $\cost(P) \ge \cost(P')$.
Since $\cost(P)$ is the MWC total weight, Eq.~\eqref{eq:wsp_solution}
gives the multiway cuts of minimal total weight, concluding the proof.
\end{proof}

%%% Local Variables:
%%% mode: latex
%%% TeX-master: "main"
%%% End:

\section{WSP used to optimize array operation fusion in Bohrium}
Stating the WSP problem formulation in a general way allows a great deal of flexibility, as long as the cost function is monotonic.
In this section, we use WSP to solve a concrete optimization problem,
demonstrating its real world use. The concrete problem is an
optimization phase within the Bohrium runtime system~\cite{kristensen2014bohrium} in which a set of array operations are
partitioned into computation kernels -- the \emph{Fusion of Array
  Operations} (FAO) problem:
\begin{definition}
  \label{def:fao_problem}
  Given a set of array operations, $A$, equipped with a strict partial
  order imposed by the data dependencies between them, $(A,\ltdep)$,
  find a partition, $P$, of $A$ for which:
  \begin{enumerate}
  \item All operations within a block in $P$ are fusible
        (Def. \ref{def:data-parallelism})
  \item For all blocks, $B\in P$, if $a_1 \ltdep a_2 \ltdep a_3$ and
        $a_1,a_3\in B$ then $a_2\in B$. (I.e.~the partition obeys dependency order).
  \item The cost of the partition (Def. \ref{def:bh_partition_cost}) is
        minimal.
  \end{enumerate}
\end{definition}
In the following, we will provide a brief description of Bohrium and show
that the WSP problem solves the FAO problem (Theorem \ref{tho:wsp_is_fao}).

\subsection{Fusion of Array Operations in Bohrium}
\newsavebox{\LstPyCodeExample}
\begin{lrbox}{\LstPyCodeExample}
\begin{lstlisting}[linewidth=0.50\linewidth,language=python, numbers=left, otherkeywords={as}]
import bohrium as bh

def synthetic():
  A = bh.zeros(4)
  B = bh.zeros(4)
  D = bh.zeros(5)
  E = bh.zeros(5)
  A += D[:-1]
  A[:] = D[:-1]
  B += E[:-1]
  B[:] = E[:-1]
  T = A * B
  bh.maximum(T, E[1:], out=D[1:])
  bh.minimum(T, D[1:], out=E[1:])
  return D
print synthetic()
\end{lstlisting}
\end{lrbox}

\newsavebox{\LstPyByteCodeExample}
\begin{lrbox}{\LstPyByteCodeExample}
\begin{lstlisting}[linewidth=0.33\linewidth, language=python, numbers=left, otherkeywords={ADD, MUL, SYNC, DEL, MAX, COPY, MIN}]
COPY A, 0
COPY B, 0
COPY D, 0
COPY E, 0
ADD A, A, D[:-1]
COPY A, D[:-1]
ADD B, B, E[:-1]
COPY B, E[:-1]
MUL T, A, B
MAX D[1:], T, E[1:]
MIN E[1:], T, D[1:]
DEL A
DEL B
DEL E
DEL T
SYNC D
DEL D
\end{lstlisting}
\end{lrbox}

\begin{figure}
    \centering
    \subfloat[][]{\label{fig:python_code_a}\usebox{\LstPyCodeExample}}\hspace{25px}
    \subfloat[][]{\label{fig:python_code_b}\usebox{\LstPyByteCodeExample}}
    \caption{A Python application that utilizes the Bohrium runtime
      system. In order to demonstrate various challenges and
      trade-offs, the application is synthetic. (a) shows the Python
      code and (b) shows the corresponding Bohrium array bytecode.}
    \label{fig:python_code}
\end{figure}

Bohrium is a computation backend for array programming languages and
libraries that supports a range of languages, such as Python, C++, and
.NET, and a range of computer architectures, such as CPU, GPU, and
clusters of these. The idea is to decouple the domain specific frontend
implementation from the hardware specific backend implementation in
order to provide a high-productivity and high-performance framework.

Similar to NumPy~\cite{van2011numpy}, a Bohrium array operation operates on a set of
inputs and produces a set of outputs~\cite{kristensen2014bohrium}.
Both input and output operands are \emph{views} of arrays. An array
view is a structured way to observe the whole or parts of an
underlying \emph{base} array. A base array is always a contiguous
one-dimensional array whereas views can have any shape, stride, and
dimensionality~\cite{kristensen2014bohrium}.  \emph{In the following, when we
  refer to an array, we mean an array view; when we refer to identical
  arrays, we mean identical array views that points to the same base
  array; and when we refer to overlapping arrays, we mean array views
  that points to some of the same elements in a common base array.}

Fig.~\ref{fig:python_code_a} shows a Python application that
uses Bohrium as a drop-in replacement for NumPy. The application
allocates and initiates four arrays (line 4-7), manipulates those
arrays through array operations (line 8-14), and prints the content of
one of the arrays (line 16).

As Bohrium is language agnostic, it translates the Python array
operations into bytecode (Fig.~\ref{fig:python_code_b}) that the
Bohrium backend can execute\footnote{For a detailed description of
  this Python-to-bytecode translation we refer to previous work
  \cite{PyHPC13_bohrium, PyHPC14_npbackend}.}. In the case of Python,
the Python array operations and the Bohrium array bytecode is almost in
one-to-one mapping. The first bytecode operand is the output
array and the remaining operands are either input arrays or input
literals. Since there is no scope in the bytecode, Bohrium uses
\texttt{DEL} to destroy arrays and \texttt{SYNC} to move array data
into the address space of the frontend language -- in this case
triggered by the Python \texttt{print} statement
(Fig.~\ref{fig:python_code_a}, line 16). There is no explicit bytecode
for constructing arrays; on first encounter, Bohrium constructs them
implicitly. %Hereafter, we use the term \emph{array bytecode} and
%\emph{array operation} interchangeable. %JA: Bliver aldrig brugt - en linie sparet! :)

In the next phase, Bohrium partitions the list of array operations
into blocks that consists of fusible array operations -- the FAO
problem. As long as the preceding constraints between the array
operations are preserved, Bohrium is free to reorder them as it sees
fit, making code optimizations based on data locality, array contraction, and
streaming possible.

In the final phase, the hardware specific backend implementation
JIT-compiles each block of array operations and executes them.

\subsubsection{Fusibility}
In order to utilize data-parallelism, Bohrium and most other array
programming languages and libraries require \emph{data-parallelism}
of array operations that are to be executed together. The property ensures that
the runtime system can calculate each output element independently
without any communication between threads or processors. In Bohrium,
all array operation must have this property.

We first introduce some operations that keep track of memory allocation,
deallocation, reads, and writes:
\begin{definition}
  Given an array operation $f$, the notation $\inop[f]$ denotes the
  set of arrays that $f$ reads; $\outop[f]$ denotes the set of arrays
  that $f$ writes; $\newop[f]$ denotes the set of new arrays that $f$
  allocates; and $\delop[f]$ denotes the set of arrays that $f$
  deletes (or de-allocates).

Furthermore, given a set of array operations, $B$, we define the following:
\begin{align*}
%    io[f]  &\equiv \outop[f] \cup \inop[f], f \in B \\
    \outop[B] &\equiv \medcup_{f\in B} \outop[f] \\
    \inop[B]  &\equiv \medcup_{f\in B} \inop[f] \\
    \newop[B] &\equiv \medcup_{f\in B} \newop[f] \\
    \delop[B] &\equiv \medcup_{f\in B} \delop[f] \\
    \extop[B] &\equiv (\inop[B]\setminus \newop[B]) \sqcup (\outop[B] \setminus \delop[B])
\end{align*}
Here, $\extop[B]$ gives the set of external data accesses.
``$\sqcup$'' is disjoint union: arrays that are both
read and written are counted twice.
\texttt{DEL} and \texttt{SYNC} are counted as having no input or output.
\end{definition}
This allows us to formulate the data-parallelism property that determines when array
operation fusion is allowed:
\begin{definition}
\label{def:data-parallelism}
A Bohrium array operation, $f$, is data parallel,
i.e.,~each output element can be calculated independently, when the following holds:
\begin{multline}
    \forall{i\in \inop[f]}, \forall{o,o'\in \outop[f]} :\\
    (i \cap o = \emptyset \lor i = o) \land (o \cap o' = \emptyset \lor o = o')
\end{multline}
In other words, if an input and an output or two output arrays
overlaps, they must be identical. %This does not apply to \texttt{DEL}
%and \texttt{SYNC} since they do not do any computation. %JA: Er det nødvendigt at pointere?
\end{definition}
Fusing array operation must preserve data-parallelism:
\begin{definition}
\label{definition:fusible}
In Bohrium, two array operations, $f$ and $f'$, are said to be fusible
when the following holds:
\begin{align}
   &\forall i' \in \inop[f'],\ \  \forall o \in \outop[f]\colon &  i' \cap o = \emptyset \lor i' = o \tag{1}\\
%    \land \notag\\
   &  \forall o' \in \outop[f'], \forall o \in \outop[f]\colon &  o' \cap o = \emptyset \lor o' = o \tag{2}\\
%    \land \notag\\
   & \forall o' \in \outop[f'], \forall i \in \inop[f]\colon &  o' \cap i = \emptyset \lor o' = i \tag{3}
\end{align}
\end{definition}
It follows from Definition \ref{def:data-parallelism} that fusible
operations are those that can be executed together without losing independent
data-parallelism.

In addition to the data-parallelism property, the current
implementation of Bohrium also requires that the length and
dimensionality of the fusible array operations are the same.

\subsubsection{Cost Model}
The motivation of fusing array operations is to reduce the overall
execution time. To accomplish this, Bohrium implements two
techniques: \bdescription
\item[Data Locality] When a kernel accesses an array multiple
  times, Bohrium will only read and/or write to that array once,
  avoiding access to main memory. Consider the two for-loops in
  Fig. \ref{lst:fuse_C_code_a} that each traverse \texttt{A} and
  \texttt{T}. Fusing the loops avoids one traversal of \texttt{A} and
  one traversal of \texttt{T}
  (Fig. \ref{lst:fuse_C_code_b}). Furthermore, the compiler can reduce
  the access to the main memory by $2N$ elements since it can keep
  the last read element of \texttt{A} and \texttt{T} in register.

\item[Array Contraction] When an array is created and destroyed within
  a single partition block, Bohrium will not allocate the array
  memory, but calculate the result in-place in one single temporary
  register variable per parallel computing thread. Consider the
  program transformation from Fig. \ref{lst:fuse_C_code_a} to
  \ref{lst:fuse_C_code_c}, in which, beside loop fusion, the temporary
  array $T$ is replaced by the scalar variable $t$. In this case, the
  transformation reduces the accessed elements with $3N$ and memory
  requirement by $N$ elements.% $N\texttt{*sizeof(double)}$ bytes.  %JA: Vi ved vel strengt taget ikke, om arrays'ne indeholder doubles?
  \edescription In
  order to utilize these optimization techniques, we introduce a WSP
  cost function that penalizes memory accesses from different
  partition blocks. For simplicity, we will not differentiate between
  reads and writes, and we will not count access to literals or register
  variables.
\begin{definition}
\label{def:bh_partition_cost}
In bohrium, the cost of a partition, $P = \{B_1,B_2,...,B_k\}$, of
array operations is given by:
\begin{equation}
%\vspace{-.3em}
    \cost(P) = \sum_{B\in P} \bytecount{\extop[B]}
%\vspace{-.2em}
\end{equation}
where the length $\bytecount{x}$ is the total number of bytes accessed
by the set of arrays in $x$.
\end{definition}
The Bohrium cost-savings when merging two partition blocks depends
only on the blocks:
\begin{prop}[Merge-savings]
    \label{prop:cost_saving}
    % Let $P = \{B_1,B_2,\ldots,\}$ and $P' = \{B_1\cup B_2,\ldots\}$,
    % i.e.~$P'$ is a successor to $P$ in the partition order and differs
    % only from $P$ by having $B_1$ and $B_2$ merged.
    Let $P$ be a partition and $P' = P \mergeop (B_1,B_2)$ be its
    successor derived by merging $B_1$ and $B_2$.
    Using the cost
    function of Def.~\ref{def:bh_partition_cost}, the difference in
    cost between the two partitions is:
    \begin{align} \label{eq:cost_saving}
      \cost(P)-\cost(P') = &\phantom{+}\,\, \bytecount{\extop[B_1] \cap \extop[B_2]} \notag\\
                           &+ \bytecount{\newop[B_1] \cap \inop[B_2]} \notag\\
                           &+ \bytecount{\outop[B_1] \cap \delop[B_2]}
    \end{align}
    Since this cost reduction depends only on $B_1$ and $B_2$, we
    define a function, $\savingop(B_1,B_2)$, that counts the savings
    from merging $B_1$ and $B_2$, which is independent of the rest of the
    partitions.
\end{prop}
\begin{proof}
  If $P = \{B_1,B_2,\ldots,\}$ and $P' = \{B_1\cup B_2,\ldots\}$, then
  the reduction in cost is
  \[
    \cost(P) - \cost(P') = \bytecount{\extop(B_1)} + \bytecount{\extop(B_2)} - \bytecount{\extop(B_1\cup B_2)}
  \]
  since all other blocks are the same. By using the fact that $B_1$
  must be executed before $B_2$, whereby
  $\inop[B_1]\cap\newop[B_2]=\emptyset$ and
  $\delop[B_1]\cap\outop[B_2]=\emptyset$, as well as the $\newop$'s
  and $\delop$'s being disjoint, direct calculation yields
  Eq.~\ref{eq:cost_saving}.
\end{proof}
Note that Prop.~\ref{prop:cost_saving} directly implies that the cost
function of Def.~\ref{def:bh_partition_cost} is positive and
monotonically decreasing, as required by Def.~\ref{def:wsp_cost}.
We next show how the problem can be formulated as a WSP instance.

\subsubsection{Constructing a WSP-problem from Bohrium bytecode}

Given a list $A$ of Bohrium array operations, a WSP problem $G =
(V,E_d,E_f)$ is constructed as follows. %JA: Vi skal nok vaelge enten f eller a som notation for array-operationer.
\begin{enumerate}
  \item The data dependencies between array operations define a partial order: $a \ltdep a'$ iff $a'$ must be executed before $a$.
  \item Each array operation $a\in A$ defines a vertex $v(a)\in V$. %JA: Skal nok skille os af med v(a) og bare bruge operationerne som knuder.
                                                                    %    For vi bruger jo lystigt blokkene som knuder senere, uden en v-wrapper.
  \item The dependency graph $E_d$ has an edge $(v(a),v(a'))$ for each pair $a,a'\in A$ with $a\ltdep a'$.
  \item The fuse-prevention graph $E_f$ has an edge $(v(a),v(a'))$ for each non-fusible pair $a,a'\in A$.
\end{enumerate}
The cost function is as in Def.~\ref{def:bh_partition_cost}, but note
that it can be calculated incrementally using Prop.~\ref{prop:cost_saving}.

\begin{figure}
 \centering
 \vspace{-0.3cm}
 \includegraphics[trim={10px 10px 10px 10px}, clip, width=\linewidth]{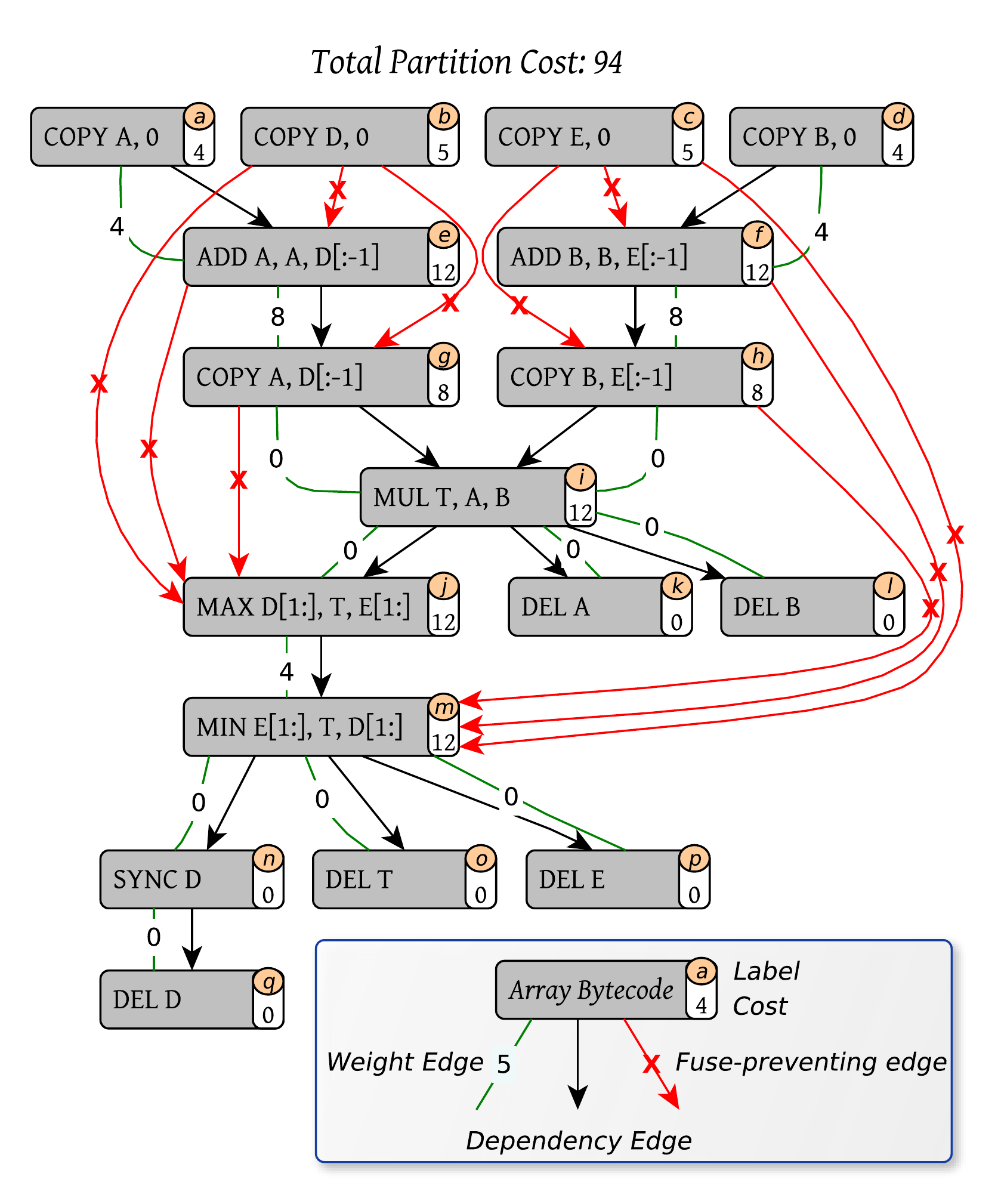}
 \caption{A partition graph of the Python application in
   Fig. \ref{fig:python_code}. For illustrative proposes, the graph
   does not include ignored weight edges \ifdefined\LongVersion
   (cf. Fig. \ref{algo:legal})\fi.}
\label{fig:dag_singleton}
\end{figure}

The complexity of this transformation is $O(V^2)$ since we may have
to check all pairs of array operations for dependecies, fusibility,
and cost-saving, all of which is $O(1)$.  Fig.~\ref{fig:dag_singleton}
shows the trivial partition, $\bot$, of the Python example, where every
array operation has its own block. The cost %of the partition %JA: Linie.
is 94.

\subsubsection{WSP solves Fusion of Array Operations}
Finally, we can show that a solution to the WSP problem also is a
solution to the FAO problem.

\begin{theorem}
\label{tho:wsp_is_fao}
  WSP solves Fusion of Array Operations.
\end{theorem}
\begin{proof}
  It is clear from Def.~\ref{def:wsp_legal_partition} and the construction above that the legal
  partitions $\hat{\Pi}_V$ are exactly all those that fulfill Properties (1) and (2) of
  Def.~\ref{def:fao_problem}. Thus, Def.~\ref{def:wsp_solution} yields a global minimum
  for all such partitions, fulfilling also Def.~\ref{def:fao_problem}(3).
\end{proof}

%%% Local Variables:
%%% mode: latex
%%% TeX-master: "main"
%%% End:

\newcommand{\AlgoLegal}{
\begin{algorithmic}[1]
\Function{Legal}{$G, e$}
    \State $(u,v)\gets e$
    \State $l\gets$ length of longest path between $u$ and $v$ in $E_d[G]$ \label{algo:legal:path}
    \If{$l = 1$} \label{algo:legal:transitive}
        \State \Return $false$
    \Else
        \State \Return $true$
    \EndIf
\EndFunction
\end{algorithmic}
}

\newcommand{\AlgoGreedy}{
\begin{algorithmic}[1]
\Function{Greedy}{$G$}
    \While{$E_w[G] \neq \emptyset$} \label{algo:greedy:while}
        \State $(u, v)\gets \Call{Find-Heaviest}{E_w[G]}$ \label{algo:greedy:heaviest}
        \If{$\Call{Legal}{G, (u,v)}$}
            \State $G\gets$ \Call{Merge}{$G,u,v$}\label{algo:greedy:merge}
        \Else
            \State Remove edge $(u,v)$ from $E_w$ \label{algo:greedy:remove}
        \EndIf
    \EndWhile
    \State \Return $G$
\EndFunction
\end{algorithmic}
}

\newcommand{\AlgoGentlyA}{
\begin{algorithmic}[1]
\Function{FindCandidate}{$G$}
    \For{$(v,u) \gets E_w[G]$}
        \If{\textbf{not} $\Call{Legal}{G,(u,v)}$}
            \State Remove edge $(u,v)$ from $E_w$
        \EndIf
    \EndFor
    \For{$(v,u) \gets E_w[G]$}
    \If{degree is less than $2$ for either $u$ or $v$ %\Statex \hspace*{\dimexpr\algorithmicindent*2}
        when only counting edges in $E_w[G]$}
            \If{$\theta[u] = \theta[v]$}
                \State \Return $(u,v)$
            \EndIf
        \EndIf
    \EndFor
    \State \Return $(\Call{NIL}{},\Call{NIL}{})$
\EndFunction
\end{algorithmic}
}

\newcommand{\AlgoGentlyB}{
\begin{algorithmic}[1]
\Function{Unintrusive}{$G$}
    \While{$(u,v) \gets \Call{FindCandidate}{G} \neq (\Call{NIL}{},\Call{NIL}{})$}
        \State $G\gets$ \Call{Merge}{$G,u,v$}
    \EndWhile
    \State \Return $G$
\EndFunction
\end{algorithmic}
}

\newcommand{\AlgoOptimalA}{
\begin{algorithmic}[1]
\Function{MergeByMask}{$G, M$}
    \State $f\gets$ \textbf{true}\Comment{Flag that indicates fusibility}
    \For{$i\gets 0$ \textbf{to} $|E_w[G]|-1$}
        \If{$M_i = 1$}
        \State $(u,v)\gets$ the $i$'th edge in $E_w[G]$
            \If{\textbf{not} \Call{Fusible}{$G, u,v$}}
                \State $f\gets$ \textbf{false}
            \EndIf
            \State $G\gets$ \Call{Merge}{$G, u, v$}
        \EndIf
    \EndFor
    \State \Return $(G, f)$
\EndFunction
\end{algorithmic}
}

\newcommand{\AlgoOptimalB}{
\begin{algorithmic}[1]
\Function{Optimal}{$G$}
    \State $G\gets$ \Call{Unintrusive}{$G$}
    \For{$(v, u) \gets |E_w[G]$}
        \If{\textbf{not} $\Call{Legal}{G, (u,v)}$}
            \State Remove edge $(u,v)$ from $E_w$
        \EndIf
    \EndFor
    \State $G_{\min}\gets$ \Call{Greedy}{$G$}\Comment{Good guess}
    \State $M_{0..|E_w[G]|}\gets 1$\Comment{Fill array $M$}
    \State $o\gets 0$\Comment{The mask offset}
    \State $Q\gets \emptyset$
    \State \Call{Enqueue}{$Q, (M, o)$}
    \While{$Q \neq \emptyset$}
        \State $(M, o)\gets$ \Call{Dequeue}{$Q$}
        \State $(G',f)\gets$ \Call{MergeByMask}{$G, M$}
        \If{$cost(G') < cost(G_{\min})$}
            \If{$f$ \textbf{and} $G'$ is acyclic}
                \State $G_{\min}\gets G'$ \Comment{New best partitioning}
            \EndIf
        \EndIf
        \For{$i\gets o$ \textbf{to} $|M|-1$}
            \State $M'\gets M$
            \State $M'_i\gets 0$
            \State \Call{Enqueue}{$Q, (M', i+1)$}
        \EndFor
    \EndWhile
    \State \Return $G_{\min}$
\EndFunction
\end{algorithmic}
}

\ifdefined\ShortVersion
\begin{figure*}
\footnotesize
\setlength{\columnsep}{5pt}
\begin{multicols}{3}
    \subfloat[][Help function for Unintrusive]{\label{algo:gently_help}\begin{tcolorbox}[width=1\linewidth]\AlgoGentlyA\end{tcolorbox}}\par
    \subfloat[][Unintrusive]{\label{algo:gently}\begin{tcolorbox}[width=1\linewidth]\AlgoGentlyB\end{tcolorbox}}\newpage \vskip -10cm
    \subfloat[][Greedy]{\label{algo:greedy}\begin{tcolorbox}[width=1\linewidth]\AlgoGreedy\end{tcolorbox}}\par
    \subfloat[][Help function for Optimal]{\label{algo:optimal_help}\begin{tcolorbox}[width=1\linewidth]\AlgoOptimalA\end{tcolorbox}}\newpage\vskip -10cm
    \subfloat[][Optimal]{\label{algo:optimal}  \begin{tcolorbox}[width=1\linewidth]\AlgoOptimalB\end{tcolorbox}}\par
    \end{multicols}
\caption{The partition algorithms where the function, $cost(G)$, returns the partition cost of the partition graph $G$.}
\end{figure*}
\fi

\section{Algorithms}

In this section, we present an exact algorithm for finding an
optimal solution to WSP (with exponential worst-case execution time),
and two fast algorithms that find approximate solutions. We
use the Python application shown in Fig.~\ref{fig:python_code} to
demonstrate the results of each partition algorithm.

\subsection{Partition graphs and chains of block merges}
All three algorithms work on data structures called {\em partition
  graphs}, defined as follows:
\begin{definition}[Partition graph]
  Given a graph $G = (V,E)$ and a partition $P$ of $V$, the
  corresponding {\em partition graph} is the graph $\pgraph{G}{P} =
  (P,\pgraph{E}{P})$ that has an edge $(B,B')$ if there is an
  edge $(u,v)\in E$ with $u\in B$ and $v\in B'$. That is, the vertices
  are the blocks, connected by the edges that cross block boundaries.
\end{definition}
\noindent
From this we build the state needed in WSP computations:
%This forms the basis of the state needed in the WSP computations:
\begin{definition}[WSP state]
  Given a WSP-instance $G = (V,E_d,E_f,\cost)$ and a partition $P$,
  the {\em WSP state} is the partition graph $\pgraph{G}{P} =
  (P,\pgraphEd{P},\pgraphEf{P})$ together with a complete weighted
  graph $\pgraph{E_w}{P}$ with weights $w(B_1,B_2) = \cost(P) -
  \cost(P \mergeop (B_1,B_2))$.
\end{definition}
Notice that $w(B_1,B_2) = \savingop(B_1,B_2)$ for the Bohrium cost
function, as shown in Prop.~\ref{prop:cost_saving}, and does not
require a full cost calculation.
\begin{definition}[Merge operator on partition graphs]
  We extend the merge operator of Def.~\ref{def:block_merge} to
  partition graphs as $\pgraph{G}{P} \mergeop (B_1,B_2) = \pgraph{G}{P
    \mergeop (B_1,B_2)}$.  This acts exactly as a vertex contraction
  on the partition graph.
\end{definition}
The merge operator is commutative in the sense that the order in a
sequence of successive vertex contractions doesn't affect the result \cite{wolle2004note}.
An auxiliary function, \Call{Merge}{}, is used in each algorithm to update the state.
\begin{definition}
  Let $S = (\hat{G},\hat{E}_w)$ be a WSP state. We define
  \[
     \Call{Merge}{(S,u,v)} = (\hat{G} \mergeop (u,v), \hat{E_w}')
  \]
  where $\hat{E_w}'$ is the updated weight graph on the edges
  incident to the new vertex $z = u\cup v$.
\end{definition}
The complexity of \Call{Merge}{} is dominated by the weight update, which requires
a $\savingop{}$ computations per edge incident to the merged vertex,
and is bounded by \bigO{V^2}. We next need a local condition for when
a merge is allowed:
\begin{lemma}[Legal merge]
  \label{lem:legal_partition_merge}
  Let $P_{1,2} = P\mergeop (B_1,B_2)$ be the successor to a legal
  partition $P\in\hat{\Pi}_V$, derived by merging blocks $B_1$ and
  $B_2$.  Then $P_{1,2}\in\hat{\Pi}_V$ if and only if
  \begin{enumerate}
  \item $(B_1,B_2)\notin \pgraphEf{P}$, and
  \item there is no path of length $\ge 2$ from $B_1$ to $B_2$ in the
        partition graph $\pgraphEd{P}$.
  \end{enumerate}
\end{lemma}
\begin{proof}
  Recall that $\hat{\Pi}_V$ is the subset of partitions in $\Pi_V$
  that satisfy Def.~\ref{def:wsp_legal_partition}.  Because $P$ is
  legal, no block contains an edge in $E_f$. Hence $P_{1,2}$ obeys
  Def.~\ref{def:wsp_legal_partition}(1) if and only if no two vertices
  $u\in B_1$ and $v\in B_2$ are connected in $E_f$, or equivalently,
  $(B_1,B_2)\notin \pgraphEf{P}$. %TODO: Forkort

  Similarly, by assumption, there are no cycles in $\pgraphEd{P}$.
  Thus, $P_{1,2}$ violates Def.~\ref{def:wsp_legal_partition}(2) if and
  only if $\pgraphEd{P}$ contains a path $B_{1}\to B' \to \cdots \to
  B_{2}$, forming the cycle $B_{1,2} \to B' \to \cdots \to B_{1,2}$ in
  $\pgraphEd{P_{1,2}}$ (where $B_{1,2} = B_1\cup B_2$).
\end{proof}

\begin{prop}[Reachability through legal merges]
  Given two legal partitions $P \ltpar P'$, there exists a successor
  chain $P \precpar P_1 \precpar P_2 \precpar \cdots \precpar P'$
  entirely contained in $\hat{\Pi}_V$, i.e.  corresponding only to
  legal block merges.
\end{prop}
\begin{proof}
  A successor chain $P \precpar P_1 \precpar \cdots \precpar P_{n-1}
  \precpar P'$ always exists in the total set of partitions $\Pi_V$,
  and all such chains are of the same length $n$.  Any such chain
  contains no partition that violates
  Def.~\ref{def:wsp_legal_partition}(1): each step is a merge,
  so once a fuse-preventing edge is placed inside a block, it would
  be included also in a block from $P'$.
  Hence we only need to worry about Def.~\ref{def:wsp_legal_partition}(2).

  We now show by induction that a successor chain consisting of only
  legal partitions exists.  First, if $P\precpar P'$, it is trivially
  so. Assume now that the statement is true for all $n \le N$, and
  consider $P\ltpar P'$ of distance $N+1$.

  %JA: TODO: Omskriv til at bruge dependency-orden isf. transitiv reduktion.
  %    Bevisskitse: Da <_d er strict findes et mindste B med $B_1 <_d B <_d B_2$ (altså: der findes ikke B' med $B_1 <_d B' <_d  B$).
  %    Sæt $P_{i+1}' = P_i \mergeop (B_1,B)$ og lad $P_{i+1}' \precpar P_{i+2}' \precpar ... \precpar P_N \precpar P'$ være en kæde af lovlige partitioner,
  %    som vi ved eksisterer da længden er $N-i < N$. Da er $P \precpar P_1 \precpar ... \precpar P_i \precpar P_{i+1}' \precpar ... \precpar P'$ en kæde
  %    af lovlige partitioner af længde N+1, bum.
  Pick any successor chain from $P$ to $P'$.  If any step violates
  Def.~\ref{def:wsp_legal_partition}(2), then let $P_{i+1} =
  P_i\mergeop (B_1,B_2)$ be the first partition in the chain that does
  so. Then there is a path $B_1\to B\to\cdots \to B_2$ in the
  transitive reduction of $\pgraphEd{P_i}$. Because $P'$ satisfies
  Def.~\ref{def:wsp_legal_partition}(2), $B_1\cup B\cup B_2$ is
  contained in a block from $P'$, whereby $P_{i+1}' \equiv P_i\mergeop
  (B_1,B) \ltpar P'$. This merge introduces no cycles, because the
  path is in the transitive reduction.  Now let $P_{i+1}' \precpar
  P_{i+2}' \precpar \cdots \precpar P'$ be a legal successor chain of
  length $N-i$, known to exist by hypothesis. Then $P\precpar
  P_1\precpar\cdots \precpar P_i \precpar P_{i+1}' \precpar \cdots
  \precpar P'$ is a length-$N+1$ successor chain consisting of only
  legal partitions, concluding the proof by induction.
\end{proof}
In particular, the optimal solutions can be reached in this way from
the bottom partition $\bot = \{\{v_1\},\{v_2\},\ldots,\{v_n\}\}$,
which we will use in the design of the algorithms.

\ifdefined\LongVersion
\begin{figure}
\footnotesize
\begin{tcolorbox}
    \AlgoLegal
\end{tcolorbox}
\caption{A help function thet determines whether the weight edge, $e \in E_w[G]$}%, should be ignored when searching for vertices to merge.}
\label{algo:legal}
\end{figure}
\fi

\ifdefined\ShortVersion
\begin{figure}
    \vspace{-0.5cm}
    \centering
    \subfloat[][Greedy]{\label{fig:dag_greedy}         \includegraphics[trim={15px 10px 12px 10px}, clip, scale=0.5, valign=t]{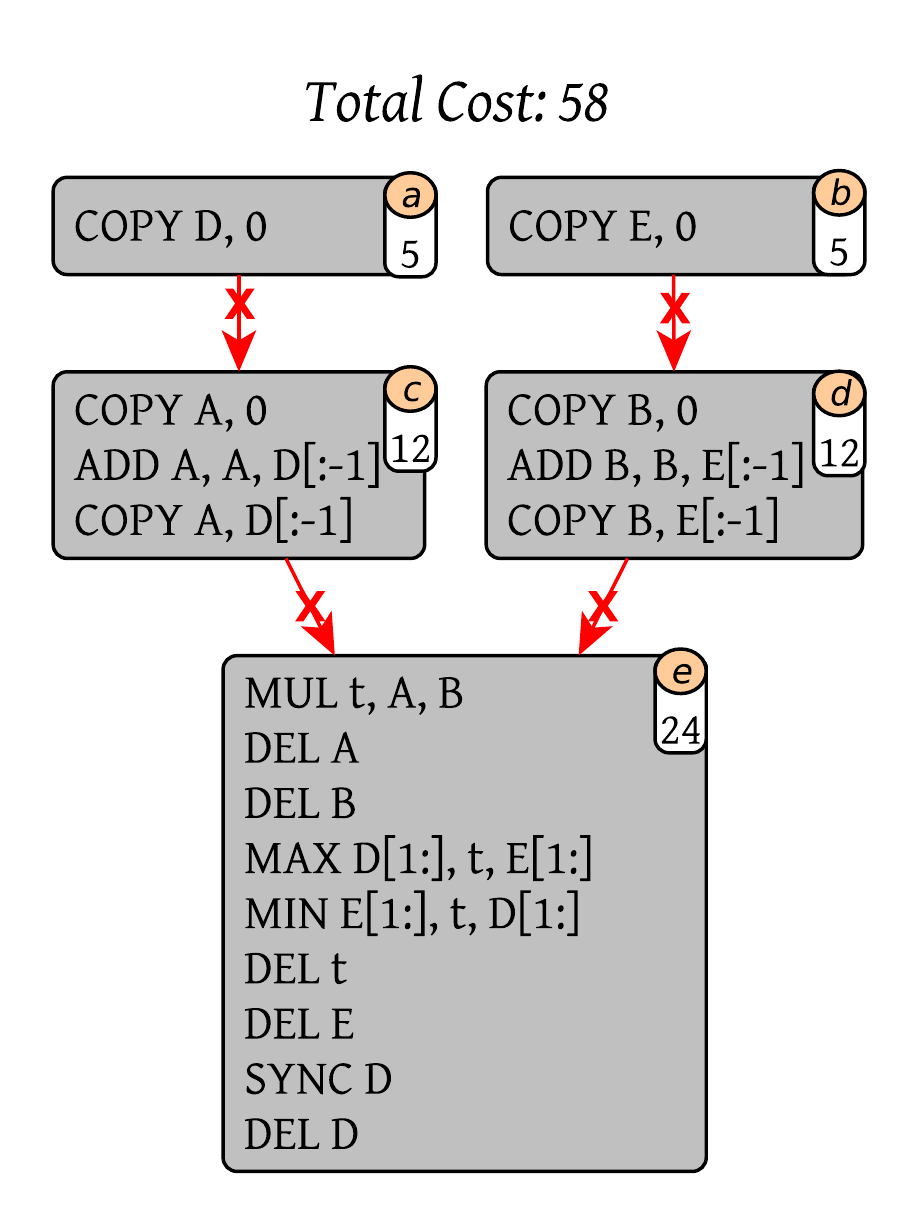}}
    \subfloat[][Unintrusive]{\label{fig:dag_gently}    \includegraphics[trim={12px 10px 15px 10px}, clip, scale=0.5, valign=t]{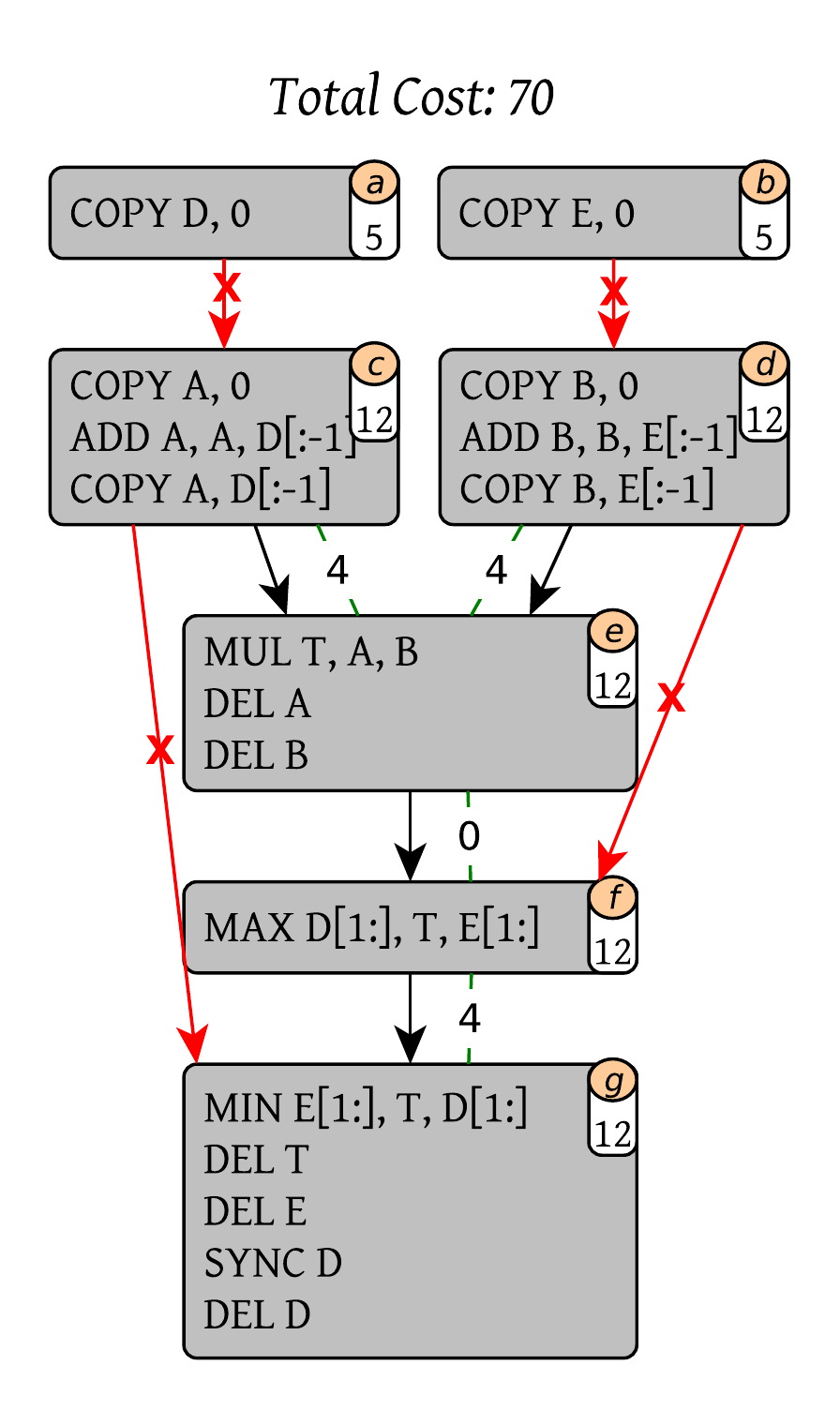}}\\
    \vspace{-0.3cm}
    \subfloat[][Optimal]{\label{fig:dag_optimal}       \includegraphics[trim={10px 10px 10px 10px}, clip, scale=0.5, valign=t]{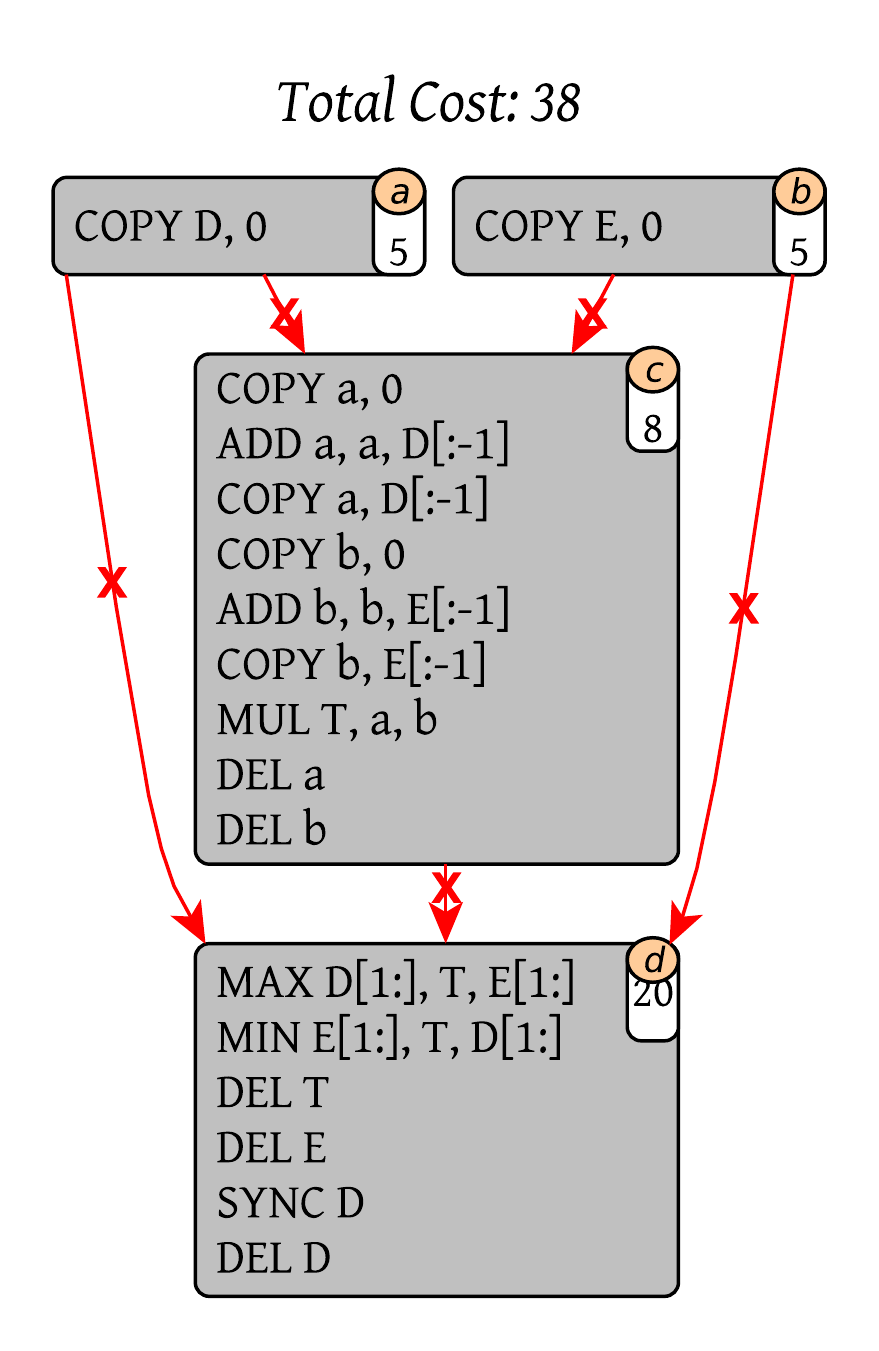}}
    \subfloat[][Linear]{\label{fig:dag_topological}\includegraphics[trim={-10px 10px -10px 10px}, clip, scale=0.5, valign=t]{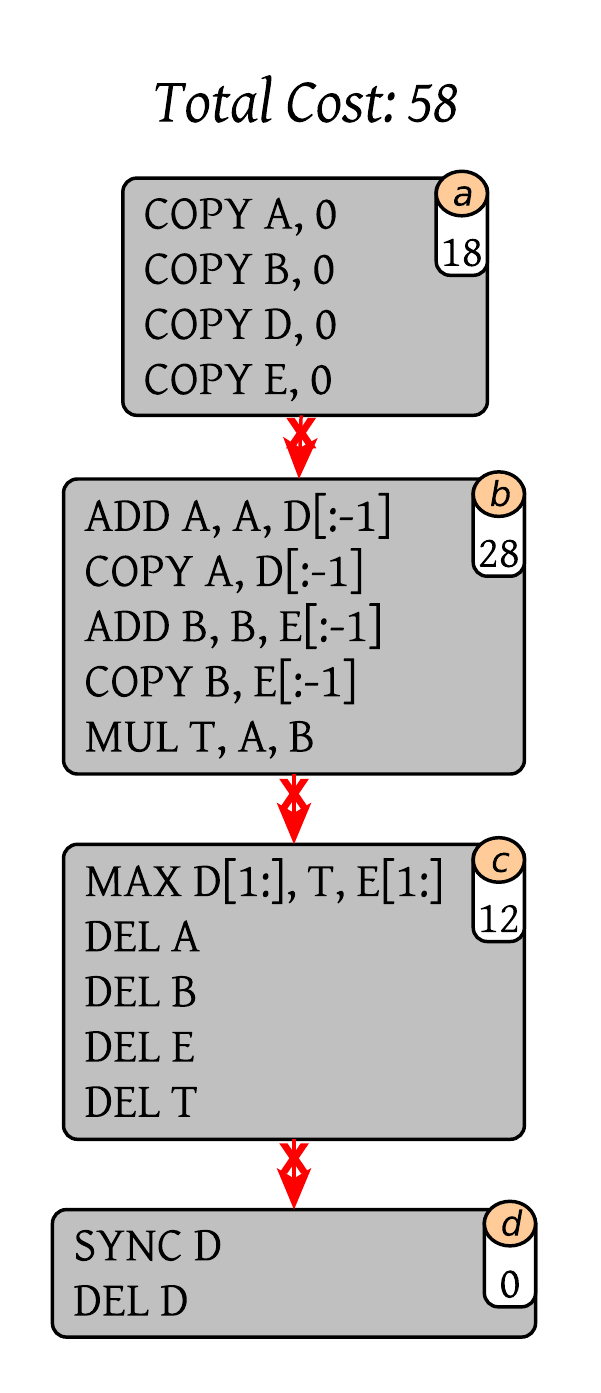}}
    \caption{Four partition graphs that are the result of running the
      four partition algorithms with Fig. \ref{fig:dag_singleton} as
      input. Lowercase variable names indicate that they are
      \emph{array contracted}.}
\end{figure}
\fi

\subsection{Unintrusive Partition Algorithm}

In order to reduce the size of the partition graph to be analyzed, we apply an
unintrusive strategy where we merge vertices that are guaranteed to be
part of an optimal solution. Consider the two vertices, $a,e$, in
Fig. \ref{fig:dag_singleton}. The only beneficial merge possibility
$a$ has is with $e$, so if $a$ is merged in the optimal solution, it
is with $e$. Now, since fusing $a,e$ will not impose any restriction
to future possible vertex merges in the graph, the two vertices are
said to be \emph{unintrusively fusible}. We formalize this property
using the {\em non-fusible sets}:
\begin{definition}[$\theta$, non-fusible set]
  The {\em non-fusible} set,  $\theta[b]$ for a block $b$
  is the set of blocks connected with $b$ in $\hat{E_d}$ through
  a path containing a non-fusible edge.
\end{definition}
\begin{theorem}
\label{tho:gentle_fusible}
Given a partition graph $\hat{G}$, let $z = u\cup v$ be the merged
vertex in $\hat{G} \mergeop (u,v)$. The vertices
$u$ and $v$ are {\em unintrusively fusible} whenever:
\begin{enumerate}
    \item $\theta[u] = \theta[v] = \theta[z]$, i.e.~the non-fusibles are unchanged. % after the merge.
    \item Either $u$ or $v$ is a pendant vertex in $\hat{E_d}$, i.e. the
         degree of either $u$ or $v$ must be $1$.
\end{enumerate}
\end{theorem}
\begin{proof}
  If Condition 1 is satisfied, any merge that is disallowed at a further
  stage due to Def.~\ref{def:wsp_legal_partition}(1) would be disallowed also without the merge.
  Similarly, merging a pendant vertex with its parent does not affect
  the possiblity of introducing cycles through future merges (Def.~\ref{def:wsp_legal_partition}(1)).
  Finally, since the cost function is monotonic, the merge cannot adversely affect a future cost.
\end{proof}

Fig.~\ref{algo:gently} shows the unintrusive partitioning algorithm. It uses
a helper function, \Call{FindCandidate}{}, to find two vertices that are
unintrusively fusible. The complexity of \Call{FindCandidate}{}
is $\bigO{E(E+V)}$, which dominates the while-loop in \Call{Unintrusive}{},
whereby the overall complexity of the unintrusive merge algorithm is
$\bigO{E^2(E+V)}$. Note that there is little need to further optimize
\Call{Unintrusive}{} since we will only use it as a preconditioner for the
optimal solution, which will dominate the computation time.

\ifdefined\LongVersion
\begin{figure}
\footnotesize
\begin{tcolorbox}
    \AlgoGentlyA
\end{tcolorbox}
\begin{tcolorbox}
    \AlgoGentlyB
\end{tcolorbox}
\caption{The unintrusive merge algorithm that only merge \emph{unintrusively fusible} vertices.}
\label{algo:gently}
\end{figure}
\fi

Fig. \ref{fig:dag_gently} shows an unintrusive partition of the Python
example with a partition cost of 70. However, the significant
improvement is the reduction of the number of weight edges in the
graph. As we shall see next, in order to find an optimal graph
partition in practical time, the number of weight edges in the graph
must be modest.

\subsection{Greedy Partition Algorithm}

Fig. \ref{algo:greedy} shows a greedy merge algorithm. It uses the
function $\Call{Find-Heaviest}{}$ to find the edge in $E_w$ with the
greatest weight and either remove it or merge over it. Note that
$\Call{Find-Heaviest}{}$ must search through $E_w$ in each iteration
since $\Call{Merge}{}$ might change the weights.

The number of iterations in the while loop (line
\ref{algo:greedy:while}) is $\bigO{E}$ since at least one weight edge is
removed in each iteration either explicitly (line
\ref{algo:greedy:remove}) or implicitly by \Call{Merge}{} (line
\ref{algo:greedy:merge}). The complexity of finding the heaviest (line
\ref{algo:greedy:heaviest}) is $\bigO{E}$, calling \Call{Legal}{} is
$\bigO{E+V}$, and calling \Call{Merge}{} is $\bigO{V^2}$ thus the overall
complexity is $\bigO{V^2E}$.

Fig. \ref{fig:dag_greedy} shows a greedy partition of the Python
example.  The partition cost is 58, which is a significant improvement
over no merge. However, it is not the optimal partitioning, as we
shall see later.

\ifdefined\LongVersion

\begin{figure}
\footnotesize
\begin{tcolorbox}
    \AlgoGreedy
\end{tcolorbox}
\caption{The greedy merge algorithm that greedily merges the vertices connected with the heaviest weight edge in $G$.}
\label{algo:greedy}
\end{figure}

\begin{figure}
 \centering
 \includegraphics[scale=0.5]{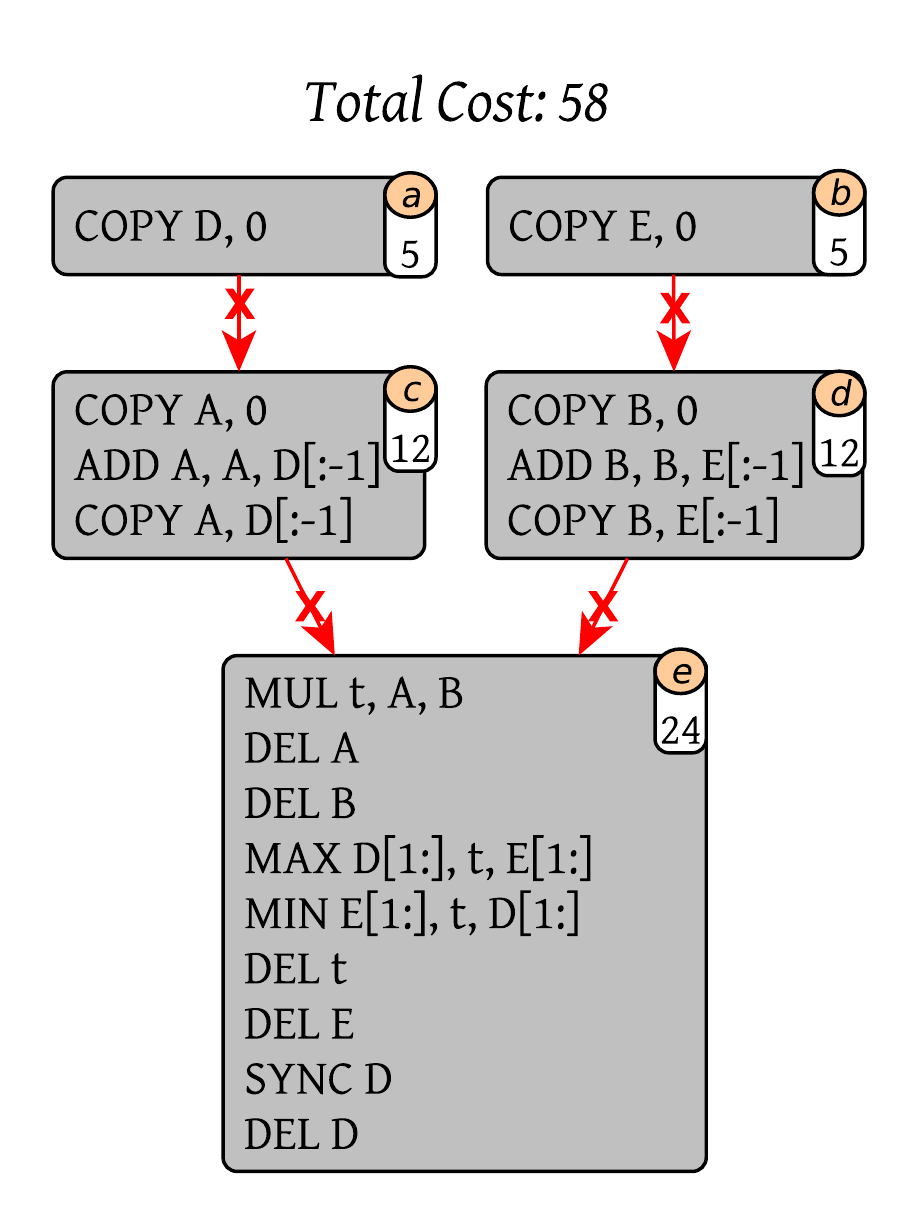}
 \caption{A partition graph of the greedy merge of the graph in Fig. \ref{fig:dag_singleton}.}
\label{fig:dag_greedy}
\end{figure}

\begin{figure}
 \centering
 \includegraphics[scale=0.5]{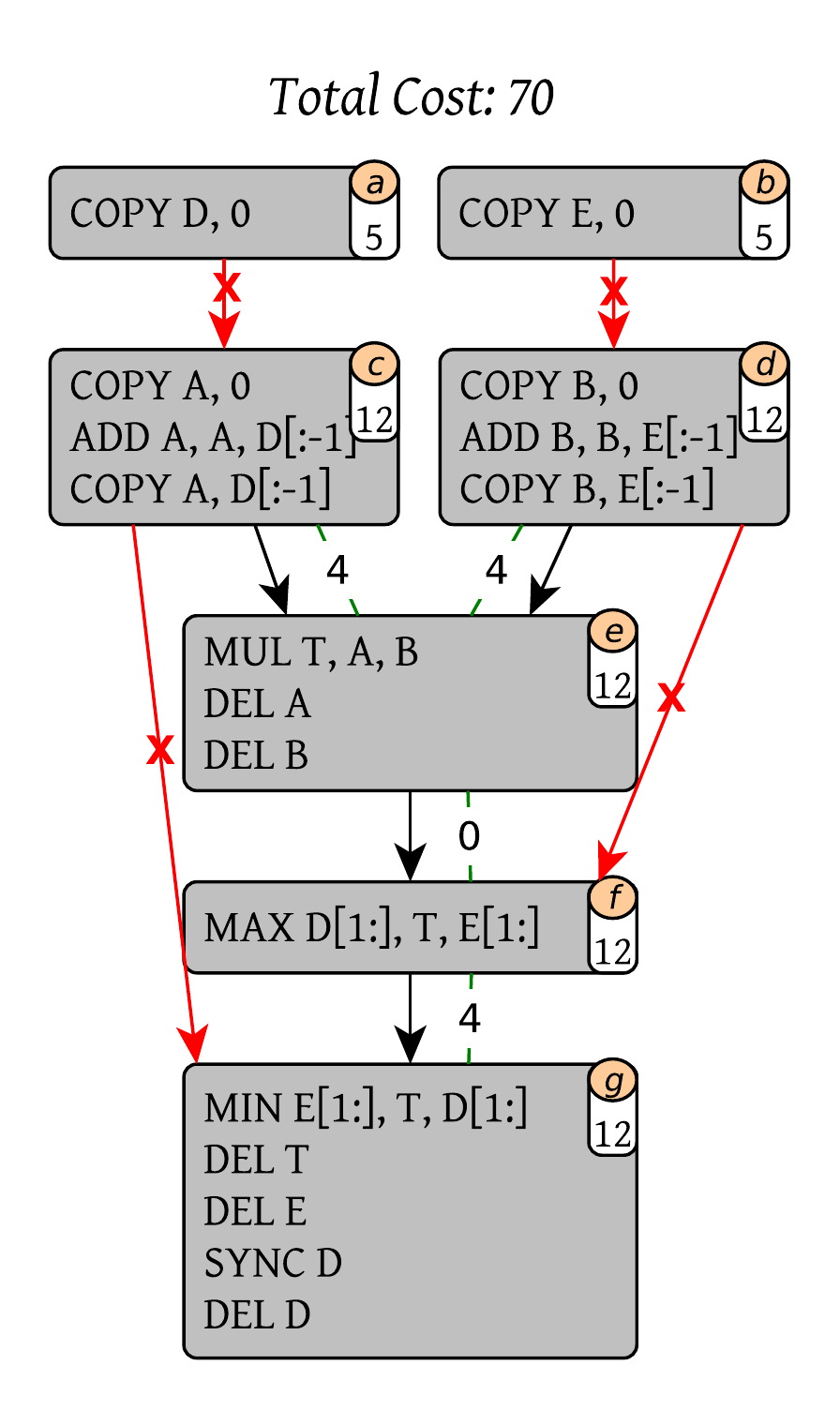}
 \caption{A partition graph of the unintrusive merge of the graph in Fig. \ref{fig:dag_singleton}.}
\label{fig:dag_gently}
\end{figure}

\fi

\subsection{Optimal Partition Algorithm}
Because the WSP problem is NP-hard, we cannot in general hope to solve
it exactly in polynomial time. However, we may be able to solve the
problems within reasonable time in common cases given a carefully
chosen search strategy through the $2^E$ possible partitions. For this
purpose, we have implemented a branch-and-bound algorithm,
exploiting the monotonicity of the partition cost (Def.~\ref{def:wsp_cost}(2)).
It is shown in Fig.~\ref{algo:optimal}, and proceeds as follows:

Before starting, the largest unintrusive partition is found. This is
the largest partition that we can ensure is included in an optimal
partition.  The blocks of the unintrusive partition will be the
vertices in our initial partition graph.  Second, a good suboptimal
solution is computed. We use the greedy algorithm for this purpose,
but any scheme will do.  We now start a search rooted in the
$\top$-partition where everything is one block. This has the lowest
cost, but will in general be illegal.  Each recursion step cuts a
weight edge that has not been considered before, if it yields a cost
that is strictly lower than the currently best partition $G_{\min}$
(if the cost is higher than for $G_{\min}$, no further splitting will
yield a better partition, and its search subtree can be ignored). If
we reach a legal partition, this will be the new best candidate, and
no further splitting will yield a better one. When the work queue is
empty, $G_{\min}$ holds an optimal solution to WSP.

\begin{figure}
 \centering
 \vspace{-10px}
 \includegraphics[width=\linewidth]{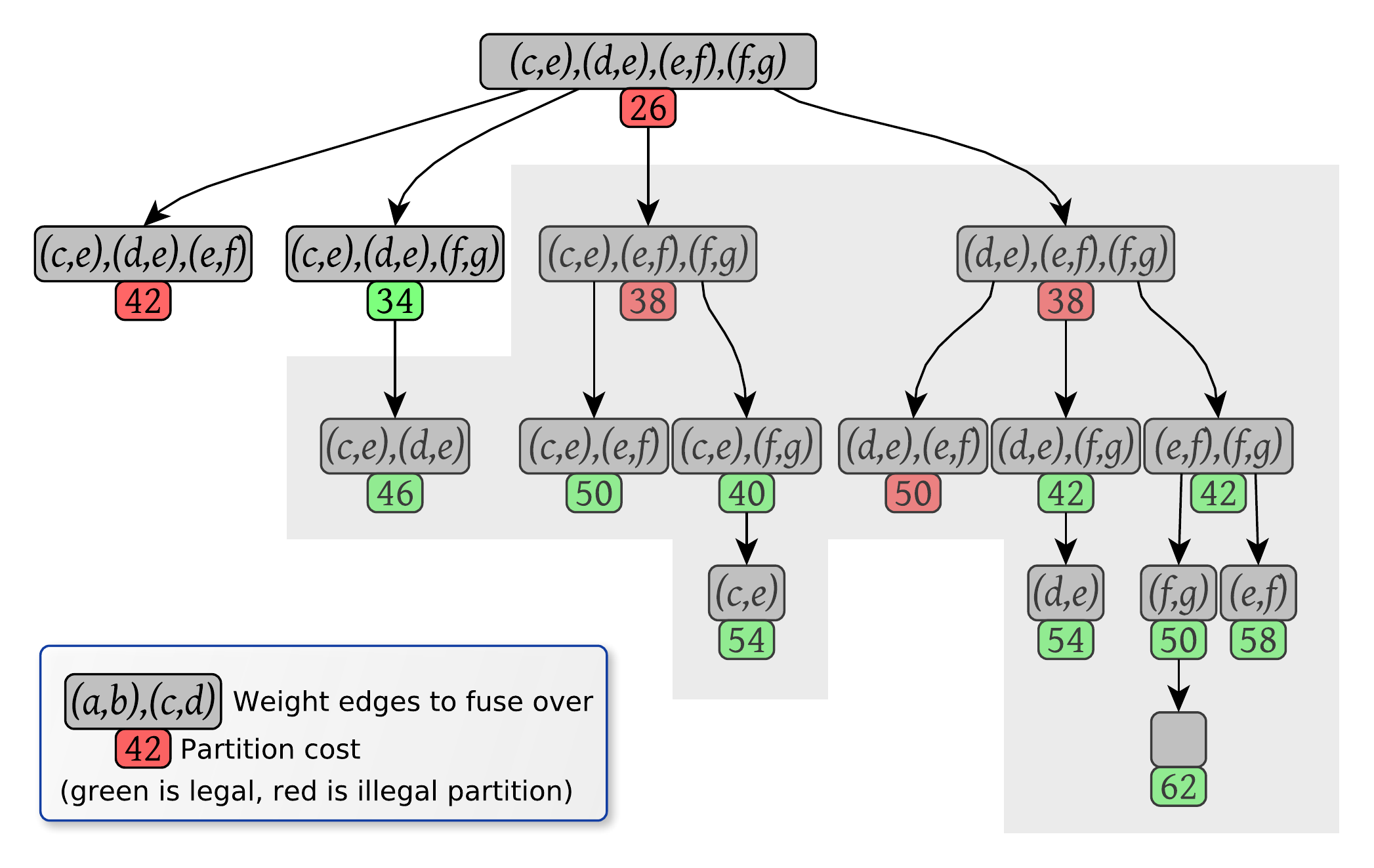}
 \caption{A branch-and-bound search tree of the unintrusively merged
   partition graph (Fig. \ref{fig:dag_gently}). Each vertex lists a
   sequences of vertex merges that build a specific graph
   partition. The grayed out area indicates the part of the search
   tree that a depth-first-search can skip because of the cost
   bound. }
 \label{fig:search_tree}
\end{figure}

Fig.~\ref{algo:optimal} shows the implementation, Fig.~\ref{fig:search_tree} shows an example of a branch-and-bound search tree, and Fig.~\ref{fig:dag_optimal} shows an optimal partition of the Python example with a partition cost of 38.

\ifdefined\LongVersion

\begin{figure}
\footnotesize
\begin{tcolorbox}
    \AlgoOptimalA
\end{tcolorbox}
\begin{tcolorbox}
    \AlgoOptimalB
\end{tcolorbox}
\caption{The optimal merge algorithm that optimally merges the
  vertices in $G$. The function, $cost(G)$, returns the partition cost
  of the partition graph $G$.}
\label{algo:optimal}
\end{figure}

\begin{figure}
 \centering
 \includegraphics[scale=0.5]{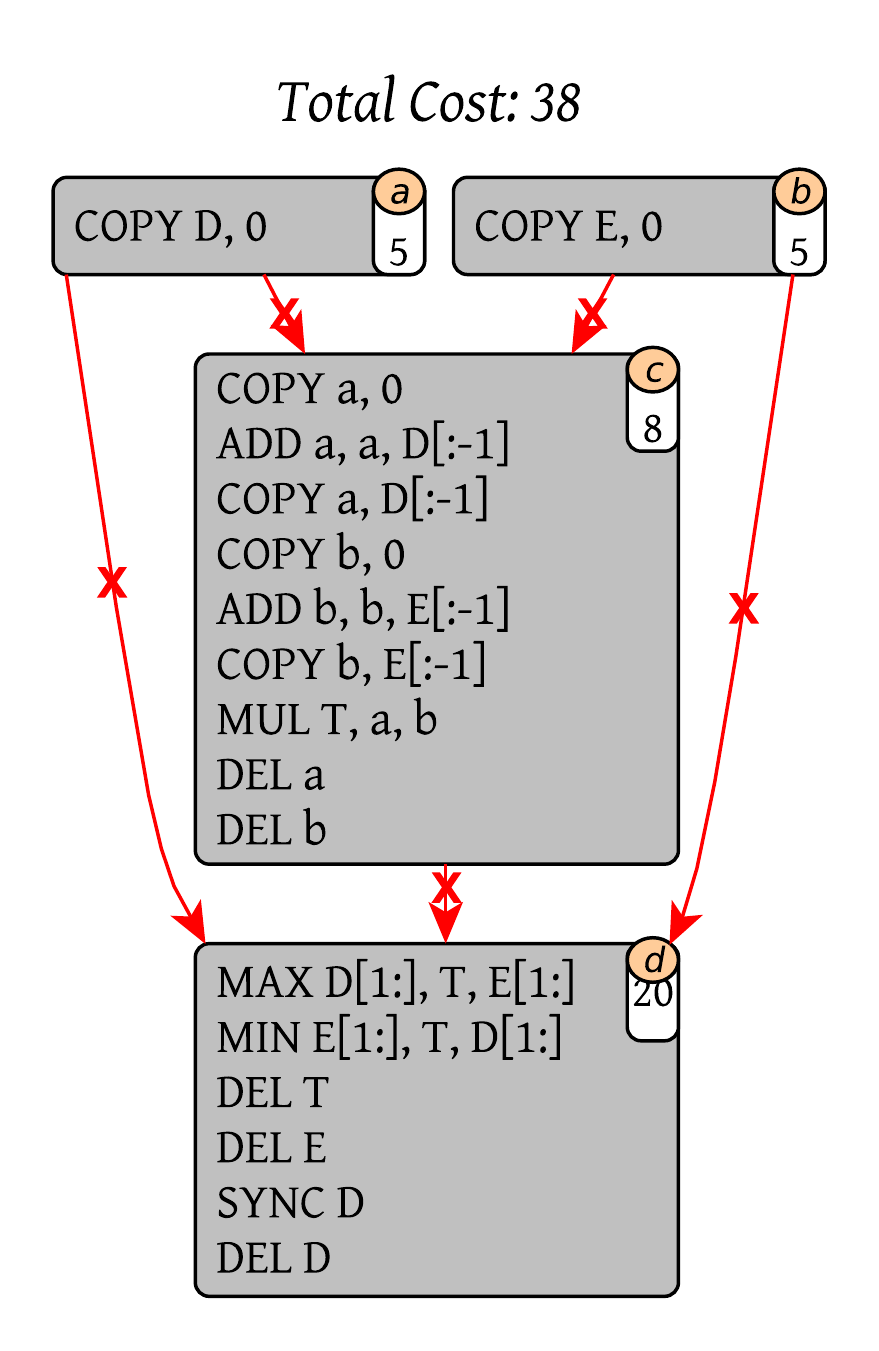}
 \caption{A partition graph of the optimal merge of the graph in
   Fig. \ref{fig:dag_singleton}.}
\label{fig:dag_optimal}
\end{figure}

\begin{figure}
 \centering
 \includegraphics[scale=0.5]{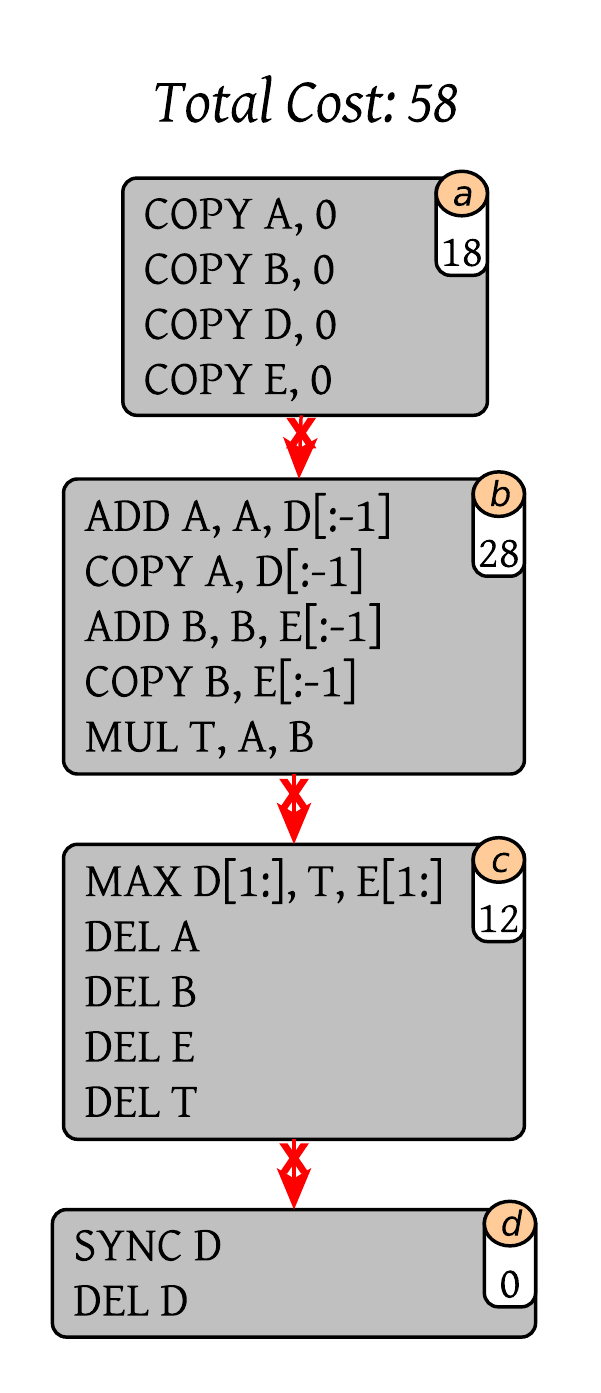}
 \caption{A partition graph of a Linear partition of the Python
   example (Fig. \ref{fig:python_code}).}
\label{fig:dag_topological}
\end{figure}

\fi

\subsection{Linear Merge}
For completeness, we also implement a partition algorithm that does
not use a graph representation. In this na\"{\i}ve approach, we simply
go through the array operation list and add each array operation to
the \emph{current} partition block unless the array operations makes
the current block illegal, in which case we add the array operation to
a new partition block, which then becomes the current one. The
asymptotic complexity of this algorithm is $\bigO{n^2}$ where $n$ is the
number of array operations.

Fig.~\ref{fig:dag_topological} show that result of partitioning the
Python example with a cost of 58.

\subsection{Merge Cache}
In order to amortize the execution time of applying the merge algorithms,
Bohrium implements a merge cache of previously found partitions of
array operation lists. It is often the case that scientific
applications use large calculation loops such that an iteration in the
loop corresponds to a list of array operations. Since the loop
contains many iterations, the cache can amortize the overall execution time
time.

%%% Local Variables:
%%% mode: latex
%%% TeX-master: "main"
%%% End:

\section{Evaluation}
In this section, we will evaluate the different partition algorithm both theoretically and practically. We execute a range of scientific Python benchmarks, which are part of an open source benchmark tool and suite named Benchpress\footnote{Available at \url{http://benchpress.bh107.org}.  For reproducibility, the exact version can be obtained from the source code repository at \url{https://github.com/bh107/benchpress.git} revision \texttt{b6e9b83}.}.

\ifdefined\ShortVersion A list of the benchmarks and the input sizes
we use throughout this evaluation can be found in the supplemental
material submitted with this paper as well as the system
specification.  \else Table \ref{tab:bench_specs} shows the specific
benchmarks that we uses and Table \ref{tab:system_specs} specifies the
host machine.  \fi When reporting execution times, we use the results
of the mean of 10 identical executions as well as error bars that
shows two standard deviations from the mean.

We would like to point out that even though we are using benchmarks
implemented in pure Python/NumPy, the performance is comparable to
traditional high-performance languages such as C and Fortran. This is
because Bohrium overloads NumPy array
operations~\cite{PyHPC14_npbackend} in order to JIT compile and execute
them in parallel seamlessly~\cite{cape:in_submission}.

\ifdefined\LongVersion
\newcommand{\Times}{{\mkern-2mu\times\mkern-2mu}}
\begin{table}
    \begin{footnotesize}
      \begin{center}
          \begin{tabular}{ l l r}
              Benchmark    & Input size (in 64bit floats) & Iterations \\
            \hline
            Black Scholes           & $1.5\Times10^6$ & $20$ \\
            Game of Life          & $10^8$ & $20$ \\
            Heat Equation         & $1.44\Times10^8$ & $20$\\
            Leibnitz PI           & $10^8$ & $20$\\
            Gauss Elimination     & $2800$ & $2799$\\
            LU Factorization      & $2800$ & $2799$\\
            Monte Carlo PI        & $10^8$ & $20$\\
            27 Point Stencil      & $4.2875\Times10^7$ & $20$\\
            Shallow Water         & $1.024\Times10^7$ & $20$\\
            Rosenbrock            & $2\Times10^8$ & $20$\\
            Successive over-relaxation & $1.44\Times10^8$ & $20$\\
            NBody                 & $6000$ & $20$\\
            NBody Nice            & $40$ plantes, $2\Times10^6$asteroids & $20$\\
            Lattice Boltzmann D3Q19 & $3.375\Times10^6$ & $20$\\
            Water-Ice Simulation  & $6.4\Times10^5$ & $20$\\
          \end{tabular}
      \end{center}
    \end{footnotesize}
  \caption{Benchmark applications}
  \label{tab:bench_specs}
\end{table}

\begin{table}
  \begin{footnotesize}
      \begin{center}
          \begin{tabular}{ l l}
            Processor:        & Intel Core i7-3770 \\
            Clock:            & 3.4 GHz            \\
            \#Cores:          & 4                  \\
            Peak performance: & 108.8 GFLOPS       \\
            L3 Cache:         & 16MB               \\
            Memory:           & 128GB DDR3         \\
            Operating system: & Ubuntu Linux 14.04.2 LTS \\
            Software:         & GCC v4.8.4, Python v2.7.6, NumPy 1.8.2\\
          \end{tabular}
      \end{center}
  \end{footnotesize}
  \caption{System specifications}
  \label{tab:system_specs}
\end{table}
\fi

\subsubsection*{Theoretical Partition Cost}

\begin{figure}
 \centering
 \includegraphics[trim={12px 10px 10px 0px}, clip, width=\linewidth]{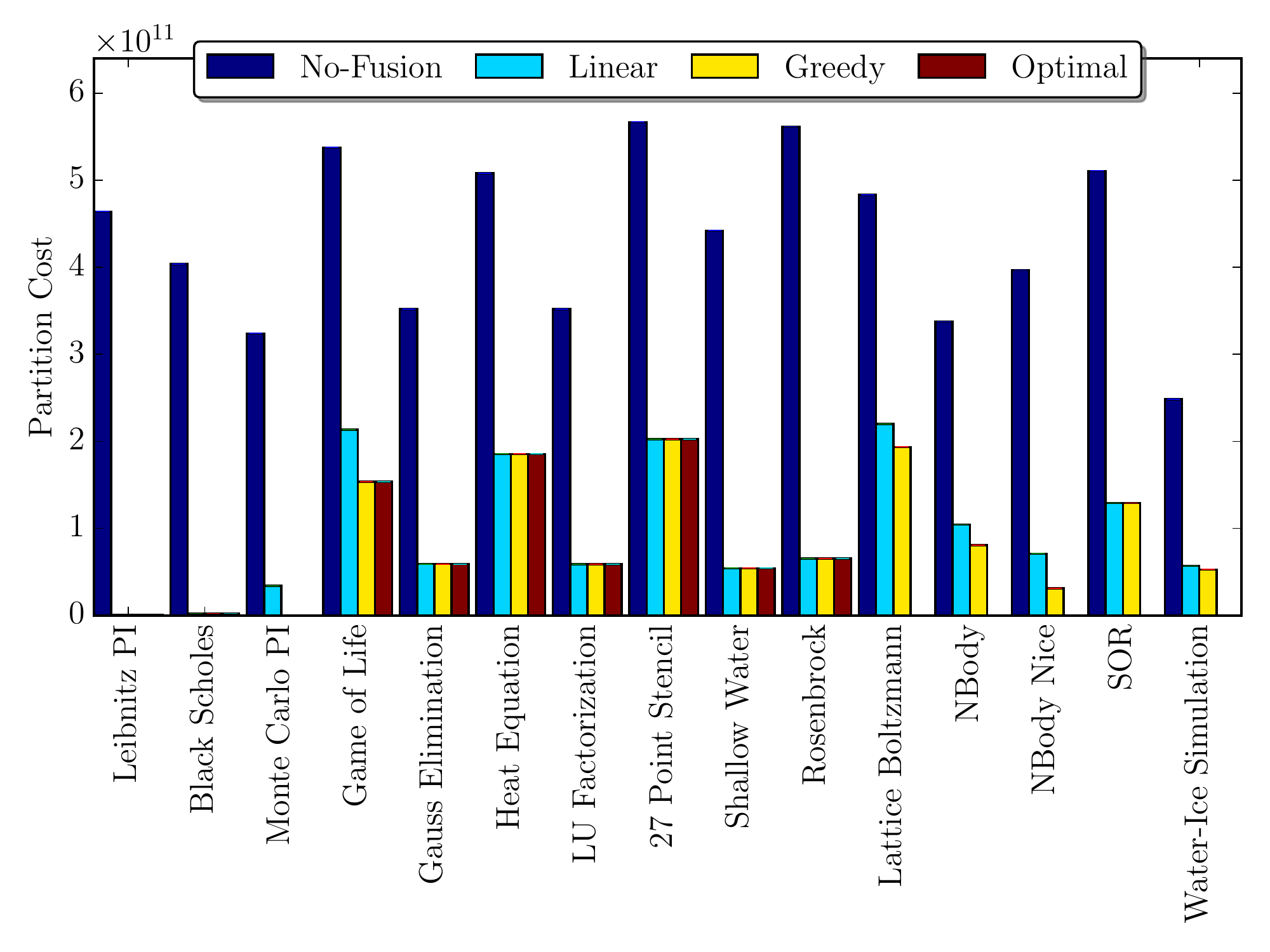}
 \caption{Theoretical cost of the different partition algorithms. NB: the last five benchmarks, Lattice Boltzmann, NBody, NBody Nice, SOR, Water-Ice Simulation, do not show an optimal solution.}
 \label{bench:unique_fuseprice}
\end{figure}

Fig. \ref{bench:unique_fuseprice} shows that theoretical partition
cost (Def. \ref{def:bh_partition_cost}) of the four different
partition algorithms previously presented. Please note that the last
five benchmarks do not show an optimal solution. This is because the
associated search trees are too large for our branch-and-bound
algorithm to solve. For example, the search tree of the Lattice
Boltzmann is $2^{664}$, which is simply too large even if the bound
can cut $99.999\%$ of the search tree away.

As expected, we observe that the three algorithms that do fusion,
Linear, Greedy, and Optimal, have a significant smaller cost than the
non-fusing algorithm Singleton. The difference between Linear and
Greedy is significant in some of the benchmarks but the difference
between greedy and optimal does almost not exist.

\subsubsection*{Practical Execution Time}
\begin{figure}
 \centering
 \includegraphics[trim={12px 10px 10px 0px},clip,width=\linewidth]{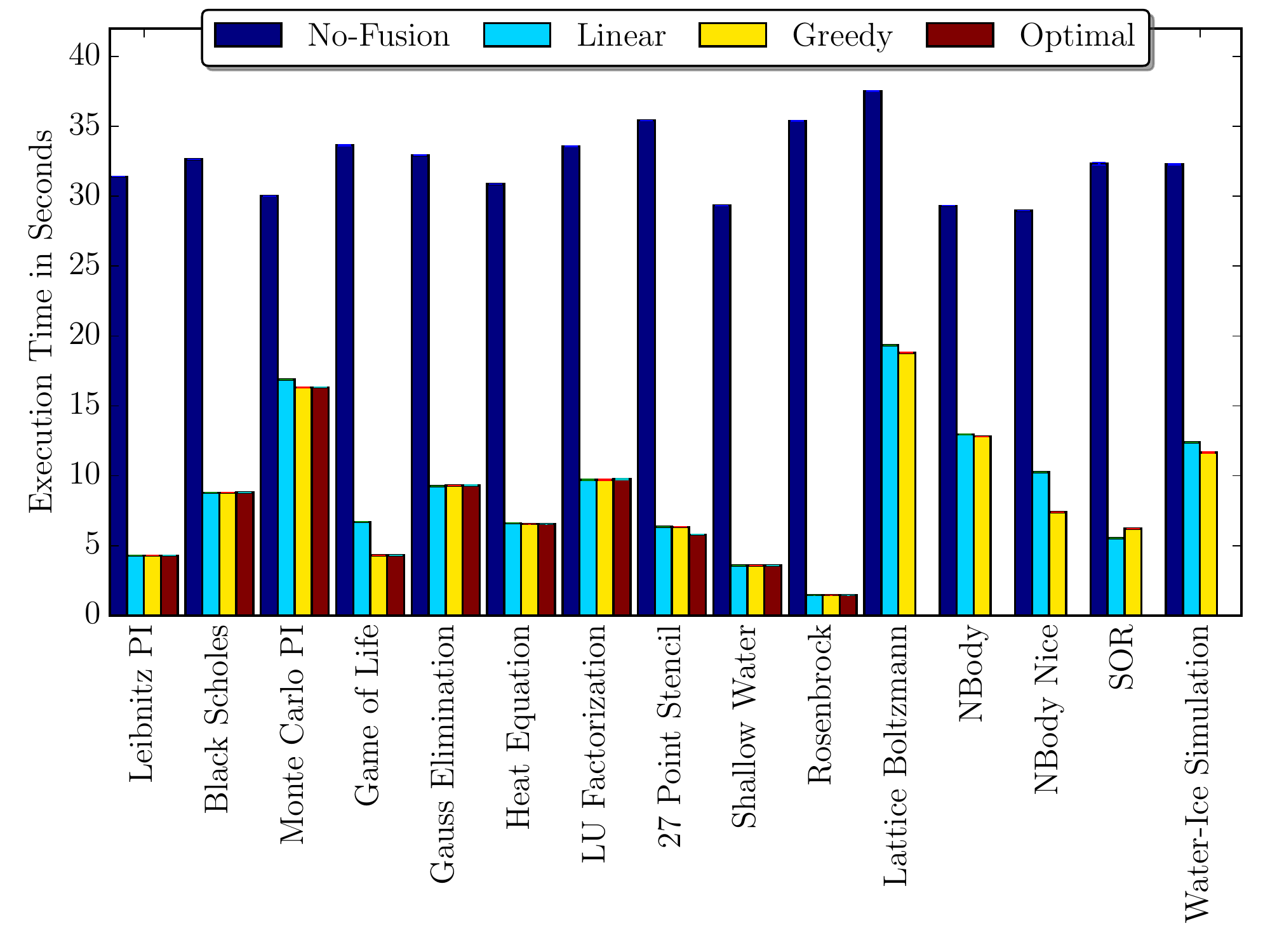}
 \caption{Execution time of the different partition algorithms using a \textbf{warm cache}.}
 \label{bench:unique_warmcache}
\end{figure}

In order to evaluate the full picture, we do three execution time
measurements: one with a warm fuse cache, one with a cold fuse cache,
and one with no fuse cache. Fig. \ref{bench:unique_warmcache} shows
the execution time when using a warm fuse cache thus we can compare the
theoretical partition cost with the practical execution time without the
overhead of running the partition algorithm. Looking at
Fig. \ref{bench:unique_fuseprice} and
Fig. \ref{bench:unique_warmcache}, it is evident that our cost model,
which is a measurement of unique array accesses
(Def. \ref{def:bh_partition_cost}), compares well to the practical
execution time result in this specific benchmark setup. However, there are
some outliers -- the Monte Carlo Pi benchmark has a theoretical
partition cost of $1$ when using the Greedy and Optimal algorithm but
has a significantly greater practical execution time. This is because the
execution becomes computation bound rather than memory bound thus a
further reduction in memory accesses does not improve performance.
Similarly, in the 27 Point Stencil benchmark the theoretical partition
cost is identical for Linear, Greedy, and Optimal, but in practice
Optimal is marginally better. This is an artifact of our cost model,
which define the cost of reads and writes identically.

\ifdefined\ShortVersion
The execution time measurements with the cold fuse cache and with the fuse cache completely disabled are in the supplemental material submitted with this paper.
\fi
With the cold fuse cache, the partition algorithm runs once in the first iteration of the computation. The results show that $20$ iterations, which most of the benchmarks uses, is enough to amortize the partition overhead\ifdefined\LongVersion
~(Fig.~\ref{bench:unique_coldcache})\fi. Whereas, if we run with no fuse cache, i.e.~we execute the partition algorithm in each iteration\ifdefined\LongVersion~(Fig.~\ref{bench:unique_nocache})\fi, the Linear partition algorithm outperforms both the Greedy and Optimal algorithm because of its smaller time complexity.

\ifdefined\LongVersion

\begin{figure}
 \centering
 \includegraphics[trim={12px 10px 10px 0px},clip,width=\linewidth]{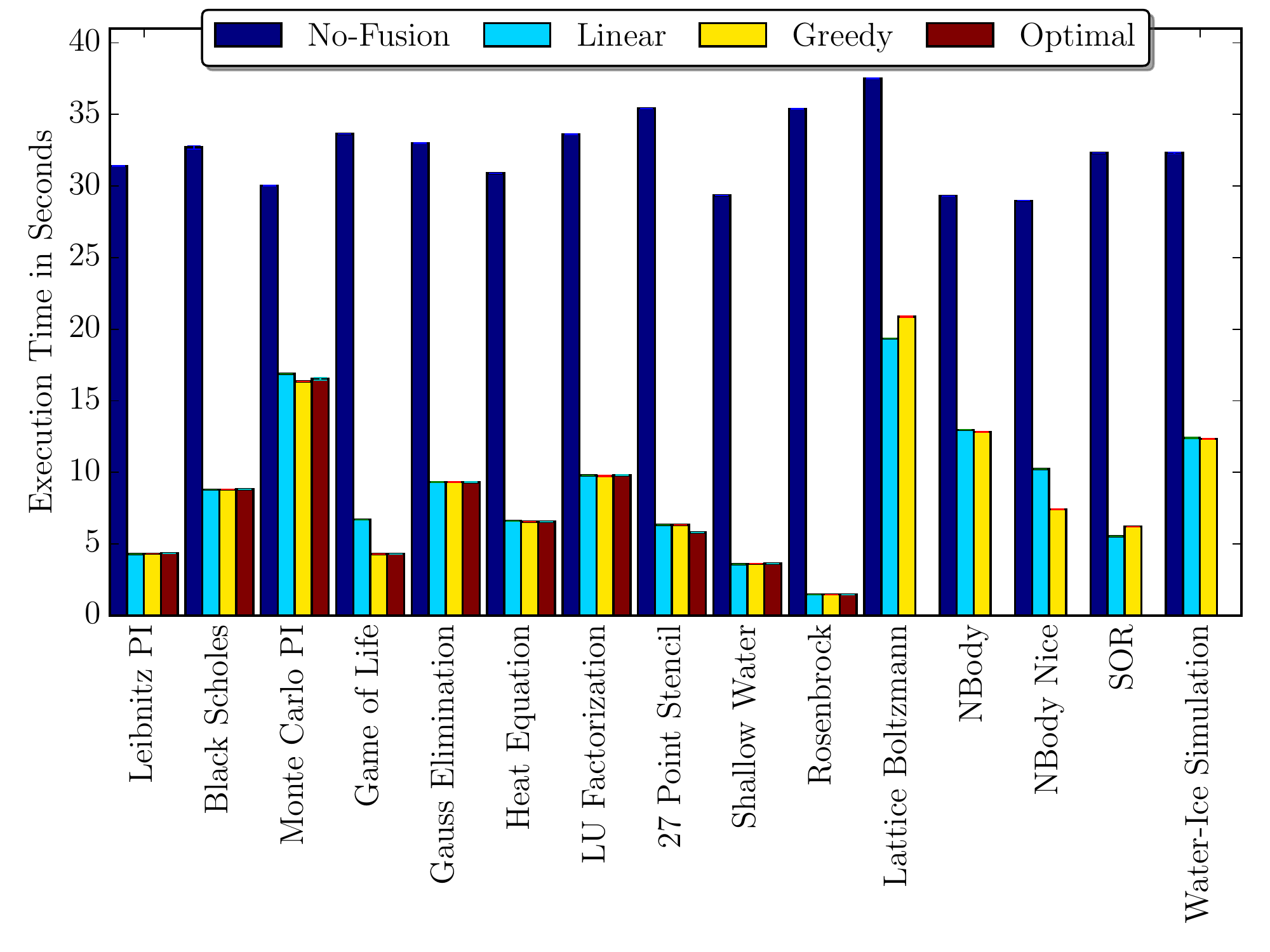}
 \caption{Execution time of the different partition algorithms using a \textbf{cold cache}.}
 \label{bench:unique_coldcache}
\end{figure}

\begin{figure}
 \centering
 \includegraphics[trim={12px 10px 10px 0px},clip,width=\linewidth]{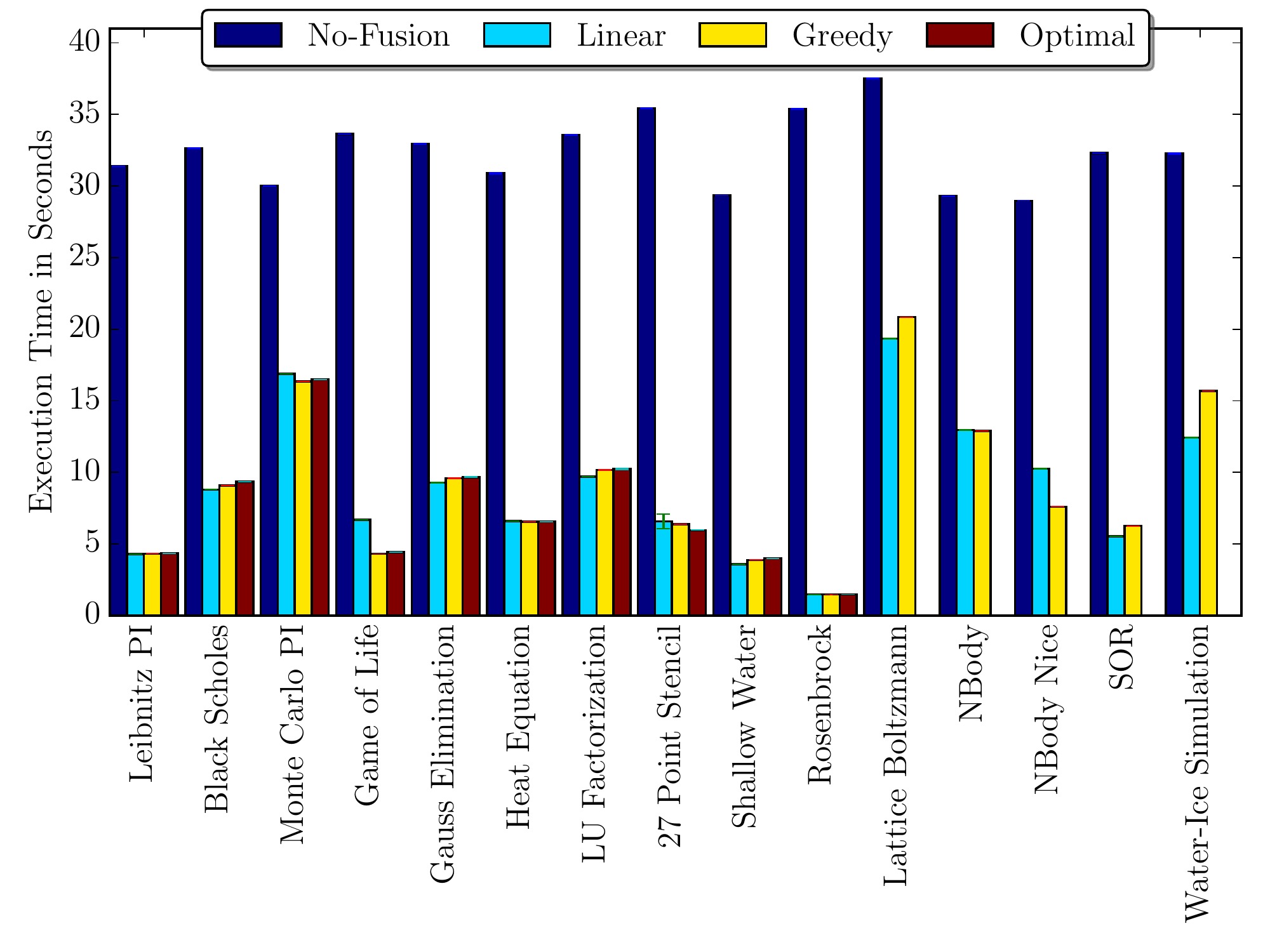}
 \caption{Execution time of the different partition algorithms using \textbf{no cache}.}
 \label{bench:unique_nocache}
\end{figure}

\fi

\subsection{Alternative Cost Model}
\label{sec:alternative-cost}
With the theoretical and practical framework we are presenting in this
paper, it is straightforward to explore the impact of alternative cost
models. In this section, we will do exactly that -- replace our cost
model with alternative cost models and evaluate the effect on the
execution time of the generated code.

Let us define and evaluate three alternative cost models, \emph{Max
  Contract}, \emph{Max Locality}, and \emph{Robinson}, which are used
in related literature~\cite{Darte2002, Megiddo1997, Robinson2014}:
\begin{definition}
\label{def:tmpelem_partition_cost}
The cost model \textbf{Max Contract} defines the cost of a partition, $P$, of array operations, $A$, as follows:
\begin{equation}
    MaxContract(A, P) = |new[A]| - \sum_{B\in P} \vbar new[B] \cap del[B] \vbar
\end{equation}
where $|new[A]|$ is the total number of allocated arrays. Thus, in this cost model, all arrays that are not contracted add $1$ to the cost.
\end{definition}
\begin{definition}
\label{def:reuse_partition_cost}
The cost model \textbf{Max Locality} defines the cost of a partition, $P$, of array operations, $A$, as follows:
\begin{equation}
    MaxLocality(A, P) = \sum_{B\in P}\sum_{f\in B}\sum_{f'\in (A\setminus B)} \vbar ext[f] \cap io[f'] \vbar
\end{equation}
In other words, this cost model penalizes each pair of array accesses
not fused with a cost of $1$.  NB: the cost is a pair-wise sum of all
identical array accesses. Thus, fusing four identical array accesses
achieves a cost saving of $6$ rather than $4$.
\end{definition}
\begin{definition}
\label{def:robin_partition_cost}
The cost model \textbf{Robinson} defines the cost of a partition, $P$, of array operations, $A$, as follows:
\begin{align}
    Robinson(A, P) &= \vbar P\vbar \notag\\
                   &+ N \cdot MaxContract(A,P)\notag\\
                   &+ N^2 \cdot MaxLocality(A,P)
\end{align}
where $N$ is the total number of accessed arrays.  In other words,
this cost model combines Max Locality, Max Contract, and penalizes the
number of partition blocks (in that priority). Furthermore, the size
of $N$ guaranties that Max Locality always attach more importance than
Max Contract which in turn always attach more importance than the
number of partition blocks.
\end{definition}

\begin{figure}
 \centering
 \includegraphics[trim={12px 10px 10px 0px},clip,width=\linewidth]{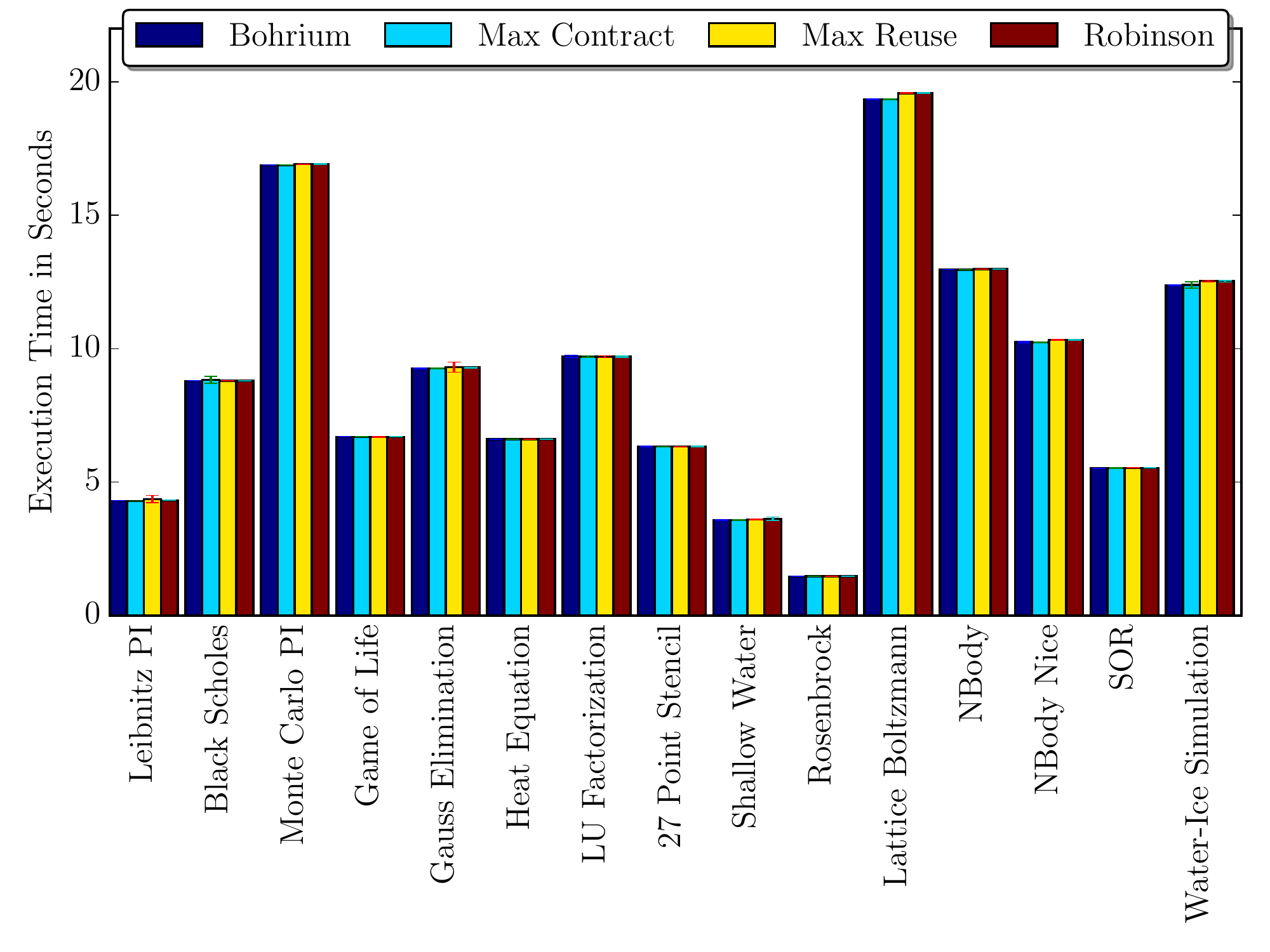}
 \caption{Execution time of the different cost models using the \textbf{Linear} partition algorithm and a \textbf{warm cache}.}
 \label{bench:all_naive_filecach}
\end{figure}

\begin{figure}
 \centering
 \includegraphics[trim={12px 10px 10px 0px},clip,width=\linewidth]{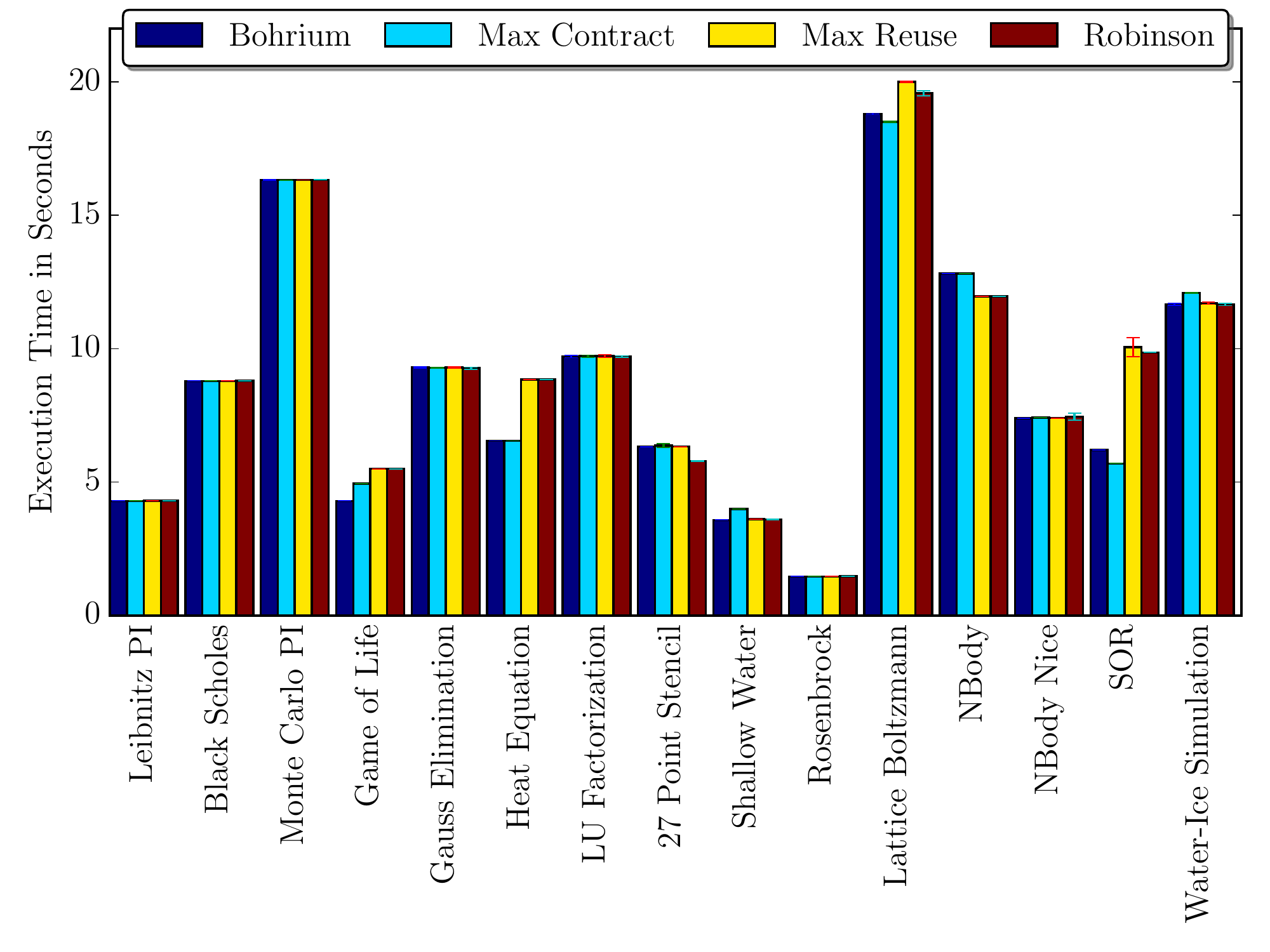}
 \caption{Execution time of the different cost models using the \textbf{Greedy} partition algorithm and a \textbf{warm cache}.}
 \label{bench:all_greedy_filecach}
\end{figure}

\begin{figure}
 \centering
 \includegraphics[trim={12px 10px 10px 0px},clip,width=\linewidth]{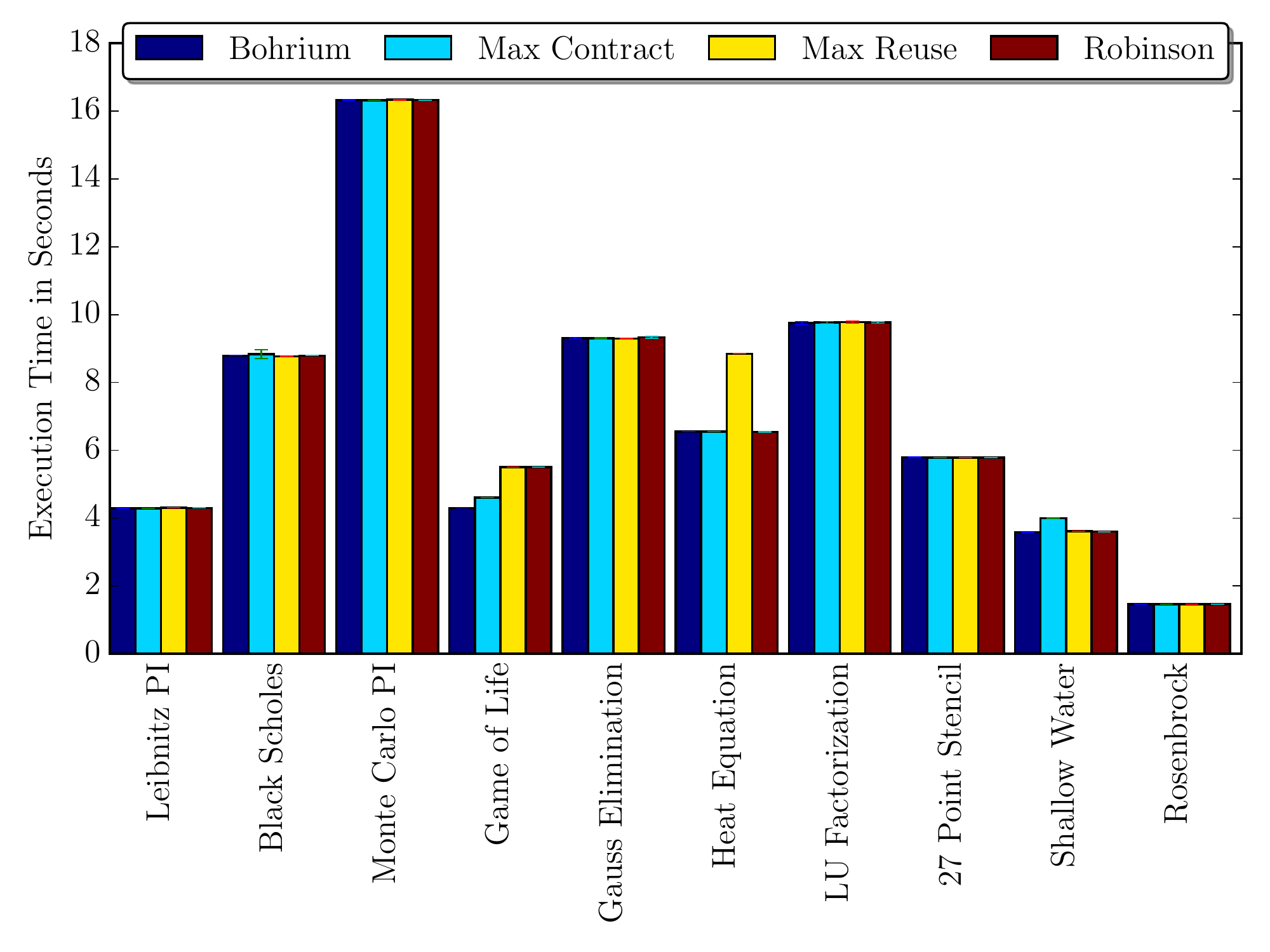}
 \caption{Execution time of the different cost models using the \textbf{Optimal} partition algorithm and a \textbf{warm cache}.}
 \label{bench:all_optimal_filecach}
\end{figure}

Fig. \ref{bench:all_naive_filecach}, \ref{bench:all_greedy_filecach},
and \ref{bench:all_optimal_filecach} compares the execution time of the cost
models using the Linear, Greedy, and Optimal partition algorithms
respectively. The execution time of the Linear algorithm is more or less
identical for all cost models.

The execution time of the Greedy algorithm shows some outliers -- in the Heat
Equation and the SOR benchmark, the performance of Max Locality and
Robinson is significantly worse than the other two.

Finally, the execution time of the Optimal algorithm shows a case, Game of
Life, where the Bohrium cost model performs better than the
others. Additionally, in the Heat Equation benchmark the performance
of Max Locality is significantly worse than the other three.

Overall, the practical performance of the four cost models is similar
for the benchmarks presented. However, there are some important
differences between them:

Since the objective of Max Contract is to maximize the number of array
contractions exclusively, there exist programs where Max Contract is
the only cost model that achieve this objective. With enough potential
data locality in a program, the other three cost models will utilize
this data locality at the expense of potential array
contractions. This was a strong motivation for Darte and
Huard~\cite{Darte2002} when they introduced an optimal solution to Max
Contract.
\ifdefined\LongVersion
Fig. \ref{lst:alain_example} shows a program fragment from \cite{Darte2002} where Max Locality fails to maximize the number of array contractions. However, in this specific program fragment, both Bohrium and Robinson obtain the same solution as Max Contract because their objective includes the maximization of the number of array contractions.
\fi

\newsavebox{\LstAlainExampleA}
\begin{lrbox}{\LstAlainExampleA}
\begin{lstlisting}[language=fortran, numbers=none, linewidth=105px, basicstyle=\scriptsize]
A(1:N)=E(0:N-1)
B = A*2 + 3
C = B + 99
D(1:N)=A(N:1:-1) + A(1:N)
E = B + C*D
F = E*4 + 2
G = E*8 - 3
H(1:N)=F(1:N)+G(1:N)*E(2:N+1)
\end{lstlisting}
\end{lrbox}

\newsavebox{\LstAlainExampleB}
\begin{lrbox}{\LstAlainExampleB}
\begin{lstlisting}[language=fortran, numbers=none, linewidth=100px, basicstyle=\scriptsize]
DO I=1,N
  A(I) = E(I-1)
ENDDO
DO I=1,N
  b = A(I)*2 + 3
  c = b + 99
  d = A(N-I+1) + A(I)
  E(I) = b + c*d
  F(I) = E(I)*4 + 2
  G(I) = E(I)*8 - 3
ENDDO
DO I=1,N
  H(I) = F(I) + G(I)*E(I+1)
ENDDO
\end{lstlisting}
\end{lrbox}

\newsavebox{\LstAlainExampleC}
\begin{lrbox}{\LstAlainExampleC}
\begin{lstlisting}[language=fortran, numbers=none, linewidth=100px, basicstyle=\scriptsize]
DO I=1,N
  A(I) = E(I-1)
ENDDO
DO I=1,N
  b = A(I)*2 + 3
  c = b + 99
  d = A(N-I+1) + A(I)
  E(I) = b + c*d
ENDDO
DO I=1,N
  f = E(I)*4 + 2
  g = E(I)*8 - 3
  H(I) = f + g*E(I+1)
ENDDO
\end{lstlisting}
\end{lrbox}

\ifdefined\LongVersion
\begin{figure}
    \centering
    \subfloat[][]{\usebox{\LstAlainExampleA}}\\
    \subfloat[][]{\usebox{\LstAlainExampleB}}\hspace{20px}
    \subfloat[][]{\usebox{\LstAlainExampleC}}\hspace{20px}
\caption{A Fortran program fragment from \cite{Darte2002}, which is based on \cite{Megiddo1997,Gao93_array_contraction}. (a) is an array operation version, (b) is the loop version that Max Locality will generate, and (c) is the loop version that Bohrium, Max Contract, and Robinson will generate.}
    \label{lst:alain_example}
\end{figure}
\fi

\section{Future Work}

The cost models we present in this paper are abstract -- they do not
take the memory architecture of the execution hardware into account.
Since the WSP formulation makes it easy to change the cost model, our
future work is to develop cost models that, in detail, model
architectures such as NUMA CPU, GPU, Intel Xeon Phi, and distributed
shared-memory machines.

Furthermore, the only requirement to the cost model in the WSP
formulation is that fusing two operations must be cost neutral or an
advantage. Thus, it is perfectly legal to have cost models that reward
fusion of specific operation types e.g. rewarding fusion of multiply
and addition instructions to utilize the FMA instruction set available
on recent Intel and AMD microprocessors.

\section{Conclusion}

In this paper, we introduce the \emph{Weighted Subroutine Partition
Problem} (WSP), which unifies program transformations for fusion of
loops, array operations, and combinators.  Contrary to previous
formulations of this problem, WSP incorporates the cost function into
the formulation, which makes WSP able to handle a wide range of
optimization objects.  Furthermore, we show that the cost function
must be part of the formulation to enable optimization objects that
minimize data locality correctly.

We prove that WSP is NP-hard and implement a branch-and-bound algorithm
that finds an optimal solution.  Out of 15 application benchmarks,
this branch-and-bound algorithm finds a solution for ten benchmarks
within reasonable execution time.

We implement a greedy algorithm that finds a \emph{good} solution to
the WSP problem, works with any cost function, and is fast enough for
Just-In-Time compilation (20 iterations is typically enough to
amortize overhead).

To evaluate various WSP algorithms, we have incorporated the
algorithms into Bohrium.  The optimization objective is then to
minimize data accesses through array contractions and data reuses
within Just-In-Time compiled computation kernels.

As expected, our evaluation shows that minimizing data accesses have a
significant performance impact.  The 15 application benchmarks
we evaluate in this paper experience a speedup ranging between 2 and
30, compared to no optimization.

However, our evaluation also shows that the various approaches to
approximate or solve the WSP have only a marginal impact on the
overall execution time of the benchmarks.  Out of 15, only one
benchmark performs significantly better using the optimal algorithm --
approximately a speedup of 1.3 compared to the greedy algorithm.
Similarly, the impact of various optimization objects is also minimal.
This tells us that approximation algorithms will give us most
of the savings, so more is won by making them faster than closer to optimal.

\bibliography{../literature}

\bibliographystyle{plain}
\end{document}